\documentclass{jfm}
\usepackage[utf8]{inputenc}
\usepackage{caption}
\usepackage{subcaption}
\usepackage{epsfig}
\usepackage{graphicx}
\usepackage{amsmath}
\usepackage[version=4]{mhchem}
\usepackage{siunitx}
\usepackage{longtable,tabularx}
\usepackage{graphicx}
\usepackage{placeins}
\usepackage{makecell}
\usepackage{multicol}
\usepackage{xcolor}

\shorttitle{The Mechanism of Lift and Pitching Moment Reversal}
\shortauthor{X. An, D. R. Williams}

\title{The Mechanism of Lift and Pitching Moment Reversal following an Actuator Burst over a Stalled Airfoil}

\author{Xuanhong An\aff{1}
  \corresp{\email{xuanhong@princeton.edu}},
  \and David R. Williams\aff{2}}

\affiliation{\aff{1} Aerospace Engineering Department,
San Jose State University, CA, 95192, USA
\aff{2}Department Mechanical, Materials and Aerospace Engineering,
Illinois Institute of Technology, IL, 60616, USA}

\begin{document}

\maketitle

\begin{abstract}
The experiment of separated flow response to a single-burst actuation over a 2-D NACA-0009 airfoil at $12^o$ angle of attack was conducted. The mechanism of the lift and pitching moment reversal following the single-burst actuation was studied. A spatially localized region of high pressure caused by a vortices-induced downwash is responsible for the lift and pitching moment reversal. Proper orthogonal decomposition (POD) of the flowfield shows that mode 2 shares a similar structure that produces the downwash and is responsible for the lift and pitching moment reversal. On the other hand, POD mode 1, which represents the direction and strength of the reverse flow on the suction side of the airfoil is responsible for the lift enhancement.
\end{abstract}

\begin{keywords}
Authors should not enter keywords on the manuscript, as these must be chosen by the author during the online submission process and will then be added during the typesetting process (see http://journals.cambridge.org/data/\linebreak[3]relatedlink/jfm-\linebreak[3]keywords.pdf for the full list)
\end{keywords}

\section{Introduction}
Flow separation (also referred to as stall) is known for degrading the performance of fluid dynamic systems such as aircraft, helicopters and wind turbines, where wings play an important role. The separated flow on the suction side of an airfoil leads to a significant lift reduction while the drag increases dramatically. For this reason, a lot of work has been carried out to explore the techniques to reattach the separated flow on a stalled wing. Since the high-speed flow often starts to detach the suction side of the airfoil in the vicinity of the leading edge, it is natural to place the actuators close to the leading edge to excite the unstable modes in the separated flow \citet{greenblatt2000control}.


In order to achieve optimal performance, leading-edge actuation (excitation) has been studied for decades for flow separation control. These studies involve a broad range of types of actuators such as synthetic jet actuators \citep{glezer2002synthetic}, plasma actuators \citep{corke2007sdbd}, combustion actuators \citep{crittenden2001combustion} and periodic blowing/suction actuators \citep{seifert1993oscillatory} \citep{williams2009lift}. For instance, the effect of periodic blowing frequency on time-averaged lift enhancement have been studied substantially \citep{DG} \citep{muller2016dynamic} \citep{raju2008dynamics}, where pneumatic actuators are used. More interestingly, \citet{glezer2002synthetic} investigated burst-mode actuation and reported that the burst-mode actuation is more efficient compared to other types of actuation in terms of time-averaged lift enhancement, because it utilizes the inherent instability in the fluid. However, despite the detailed flowfield analysis, these studies focus on the actuators' performance in terms of the time-averaged quantities (e.g. time-averaged lift).

In the cases where we want to actively control the flow separation \citep{DW} in a time-varying manner (e.g. unsteady flow separation), the dynamic characteristics of the actuators become rather important. In fact, \citet{kerstens2011closed} suggested that the bandwidth of the unsteady flow separation control is primarily limited by the time response of the actuators. Here, the time response is characterized by the time delay of the aerodynamic loads (e.g. lift) variation in response to particular input (e.g. step input) \citep{cattafesta2011actuators}. Therefore, studying the actuation delay is important for controlling the unsteady flow separation including actuator design. 

In recent studies, \citet{an2016modeling} and \citet{an2020hybrid} successfully modeled the actuator delay with low-order models on a stalled wing when a time-varying leading-edge actuation is used to control the separated flow. However, in order to improve the actuator design, being able to identify the low-order mathematical representation of the actuator delay is not enough, more comprehensive studies on the flow physics behind the actuation delay are needed.  

The dynamic response of separated flow to leading-edge actuation has been carried out by several scholars. \citet{amitay2006flow} investigated the time evolution of separated flow following a step input modulated on a sequence of burst signals. They found that the lift declines initially following the rising edge of the step input prior to the lift enhancement taking over. Later on, they \citep{amitay2006flow} conducted a similar experiment using impulse-like leading-edge actuation and found that the initial decline of the lift behaves similarly compared to that in the step input case. The initial decline of the lift is referred to as 'lift reversal' by \citet{williams2018alleviating} and they suggested that this minimum phase behavior produces a right half-plane zero that limits control bandwidth \citep{10.5555/1121635}. Given the fact that the lift reversal in the step input cases is the same as it is in the impulse input cases, investigating the impulse input is sufficient to understand the mechanism of the lift reversal. On the other hand, \citep{an2017response} also found a pitching moment reversal, which behaves very similar to the lift reversal following a burst mode actuation. 

In addition, \citet{williams2018alleviating} reported that in flow separation control with burst-mode actuation, the inherent time delay (lift reversal) is due to the nature of the separated flow and independent of the actuators. In their work, by comparing the lift response to the actuation from different actuators including zero-net-mass-flux (ZNMF) actuators, combustion actuators, and Lorentz force actuators, they found very similar lift reversal following an impulse actuation regardless of the type of actuators. Meanwhile, a similar left reversal is also observed by \citet{zong2018airfoil} using plasma actuators. 

In this paper, we intend to study the mechanism of lift and pitching moment reversal following an impulse (single-burst) input to ZNMF actuators over a NACA-0009 airfoil at $12^o$ angle of attack. It is important to point out that first, it has been mentioned earlier that the lift and pitching moment reversal following impulse actuation shares the same trend and mechanism as it is in the step input cases, second, the lift and pitching moment reversal is independent of the types of actuators as long as the actuation is meant to excite the instability of the separated flow. Therefore, the findings in this paper also apply to step (as well as other time-varying) input cases and other types of actuators (e.g. combustion actuators, plasma actuators).

The remaining of this paper is arranged as follows. A detailed description of the experimental setup is given in section \ref{sec:exp}. Section \ref{Sec:single} briefly discusses the flowfield evolution associated with the lift and pitching moment variation following the single-burst actuation. In section \ref{sec:reversal} a study of the mechanism behind the lift and pitching moment reversal is carried out using two methods. In the first method, we investigate the lift and pitching moment reversal by connecting the flowfield with its associated lift, pitching moment and pressure measurements. In the second method, we perform Proper Orthogonal Decomposition (POD) on the flowfield to identify the flow structure that is responsible for the lift and pitching moment reversal. The conclusion is given in section \ref{Sec:conc}. 







\section{Experimental Setup}\label{sec:exp}

\begin{figure}
\centering
\includegraphics[width=.7\textwidth]{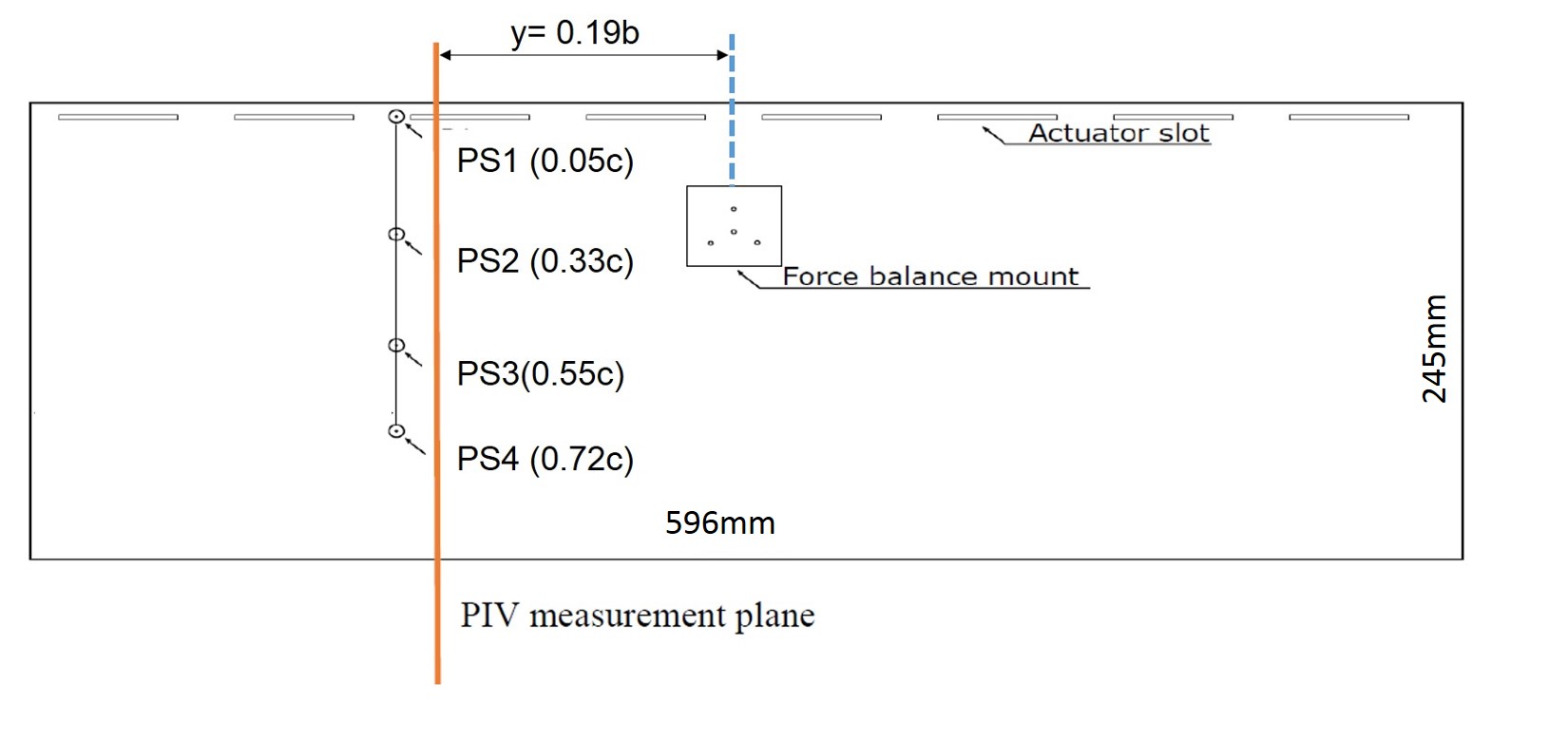}
\caption{Plan view of the wing with NACA 0009 profile. Pressure sensor, force balance, and actuator locations are shown.}
\label{fig:wing}
\end{figure}

The experiments were conducted in the Andrew Fejer Unsteady Flow Wind Tunnel at Illinois Institute of Technology. The wind tunnel has cross-section dimensions $600mm \times 600mm$. The right-hand coordinate system is defined with the origin at the leading edge of the wing and the x-axis in the flow direction, the y-axis pointing upward, and the z-axis pointing in the direction of the left side of the wing. A nominally two-dimensional NACA0009 wing with a wingspan $b = 596mm$ and chord length $c = 245mm$ was used as the test article that is shown in figure \ref{fig:wing}. The freestream speed was $U_{\infty}=3m/s$, corresponding to a convective time $t_{convect}=c/U_{\infty}=0.082 s$.  Dimensionless time $t^+$ is normalized by the convective time so that $t^+=t/t_{convect}$.  The chord-based Reynolds number is $Re_c = 49,000$. The angle of attack of the wing was fixed at $\alpha=12^o$, at which the flow is fully separated on the suction side of the airfoil. In the remaining of this paper, the lift, $L$ is nondimensionalized as the lift coefficient, $C_L = \frac{L}{0.5\rho U_{\infty}^2cb}$, where $\rho$ is the air density, the pitching moment $M$ is nondimensionalized as the pitching moment coefficient, $C_M = \frac{M}{0.5\rho U_{\infty}^2c^2b}$, and the pressure $P$ is nondimensionalized as the pressure coefficient, $C_P = \frac{P}{0.5\rho U_{\infty}^2}$.

Eight piezoelectric (zero net mass) actuators were installed in the leading edge of the wing. The slots of the actuators were located 0.05c from the leading edge with an exit angle of 30 degrees from the tangent to the surface on the suction side of the wing. The dimension of each actuator orifice slot is $2mm \times 40mm$. A plan view of the wing is given in figure \ref{fig:wing}.   

Surface pressure measurements were made with All-Sensors D2-P4V Mini transducers built into four chord-wise locations on the wing.  The pressure range for these sensors is +/- 1 inch of water. The four pressure sensors are shown in figure \ref{fig:wing} as PS1 – PS4, and the corresponding pressure coefficient measurements will be denoted as $C_P 1$ - $C_P 4$ in the rest of this paper.
Forces and moments were measured with an ATI, Inc. model Nano-17 force balance located inside the model at $30\%$ of the chord, which is also the center of gravity of the wing. The reference point of the pitching moment is located at $25\%$ of the chord. 

\begin{figure}
\centering
\includegraphics[width=.8\textwidth]{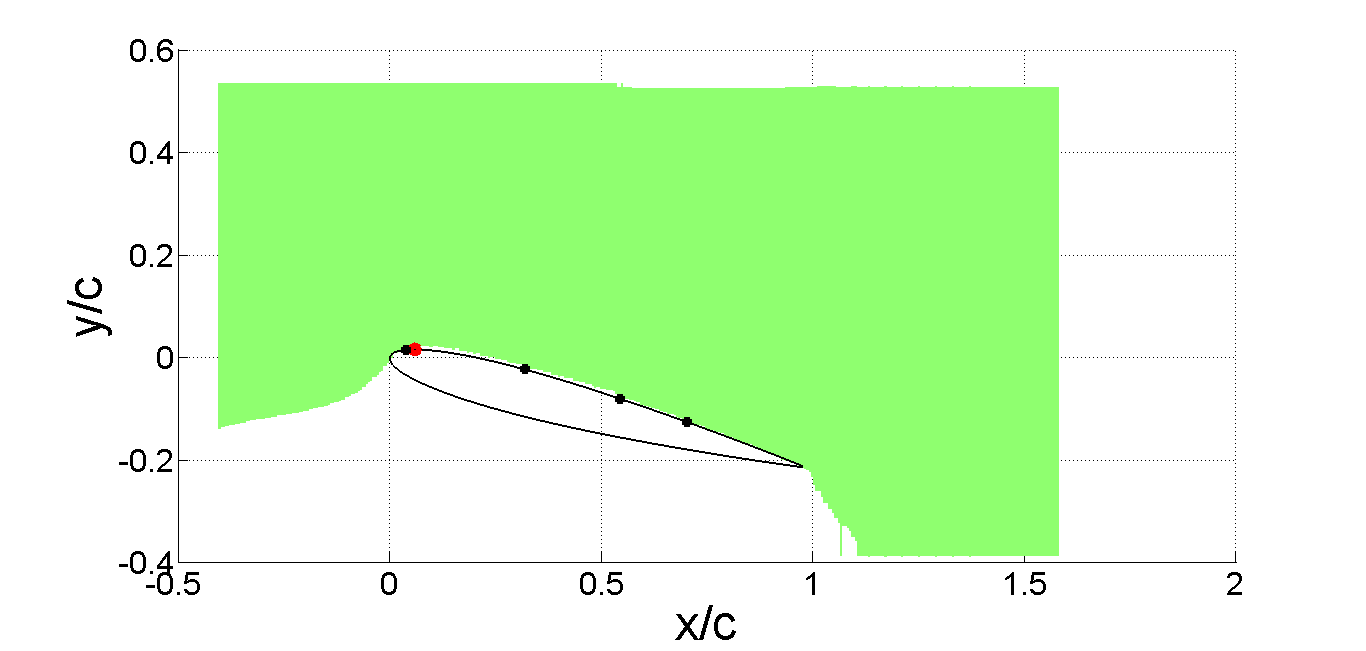}
\caption{Side view of the wing and the PIV measurement window.  Pressure sensor locations are shown by the black dots, and the actuator locations are shown by the red dot. }
\label{fig:PIV_window}
\end{figure}

Flowfield measurements using Particle Image Velocimetry (PIV) were obtained in the x,y plane located at z=0.19b away (indicated by the orange line in figure \ref{fig:wing}) from the centerline. The PIV data window in the x,y plane is shown in figure \ref{fig:PIV_window} with green color.  The small red circle denotes the streamwise location of the actuators and the black dots are the locations of the pressure sensors. The time interval between the phase-averaged PIV measurements is $0.005s$  $(0.0625t^+)$, which resulted in 800 phases covering 4s $(50t^+)$. The phase averaging was done by averaging 100 flow field images for each phase.  The initial (reference) phase corresponded to the beginning of the actuator burst signal.

\begin{figure}
	\centering
	
	        \includegraphics[width=3.0in]{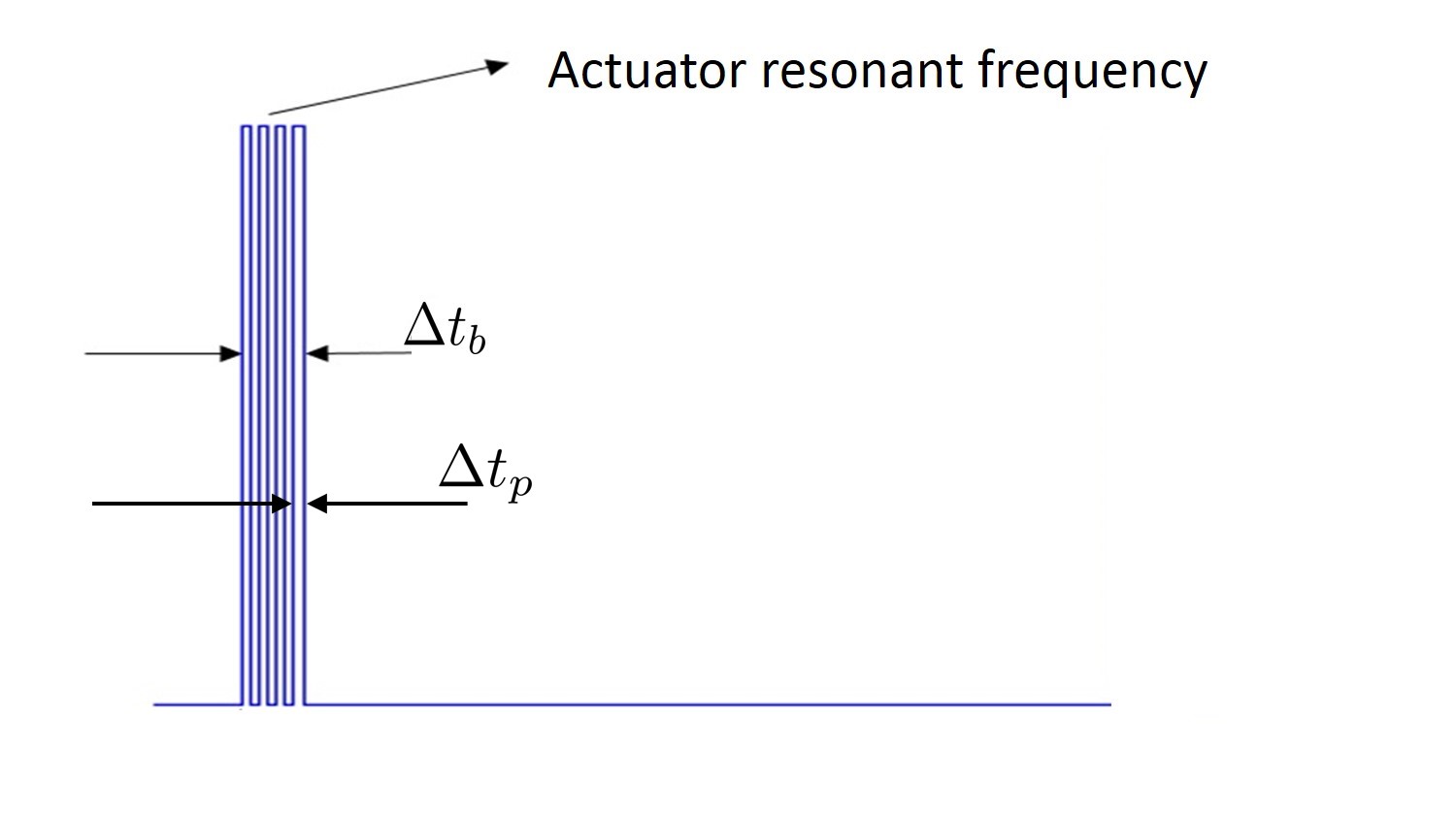}
	          \caption{Single-burst actuation consists of four$\Delta t_p$ pulses.}
	          \label{fig:input_single}

    \caption{Input signal to the actuators.}
    \label{fig:input_sig}		
\end{figure}

\FloatBarrier
To produce the maximum exit jet velocity, the ZNMF actuators are operated at their mechanical resonance frequency, $f_r=400Hz$ with a pulse width of $\Delta t_p=0.03125t^+$, and 60 Volts ($V$) amplitude. A second square wave signal was superposed on the $400Hz$ carrier signal to create the `burst signal'.  The burst signal width is $\Delta t_b=0.125t^+$. Therefore, the actual input signal to the actuators is a short burst signal containing 4 high-frequency ($400Hz$) pulses, the amplitude of the input signal, A$=60V$ for all the cases in the current research (figure \ref{fig:input_sig}). The  peak exit jet velocity measured with a hot-wire anemometer at the actuator exit is $4.9m/s$ corresponding to the peak momentum coefficient, $C_{\mu}=\frac{\rho {V_{jet}}^2 A_{jet}}{0.5\rho U_{\infty}^2 cb}=0.01$, where $V_{jet}$ is the peak velocity of the actuation jet, $A_{jet}$ is the opening area of the actuators.

\FloatBarrier

\section{Single-burst Actuation}\label{Sec:single}	
\begin{figure}
	\centering
    \begin{subfigure}{0.3\textwidth}
    	        \includegraphics[width=1.7in]{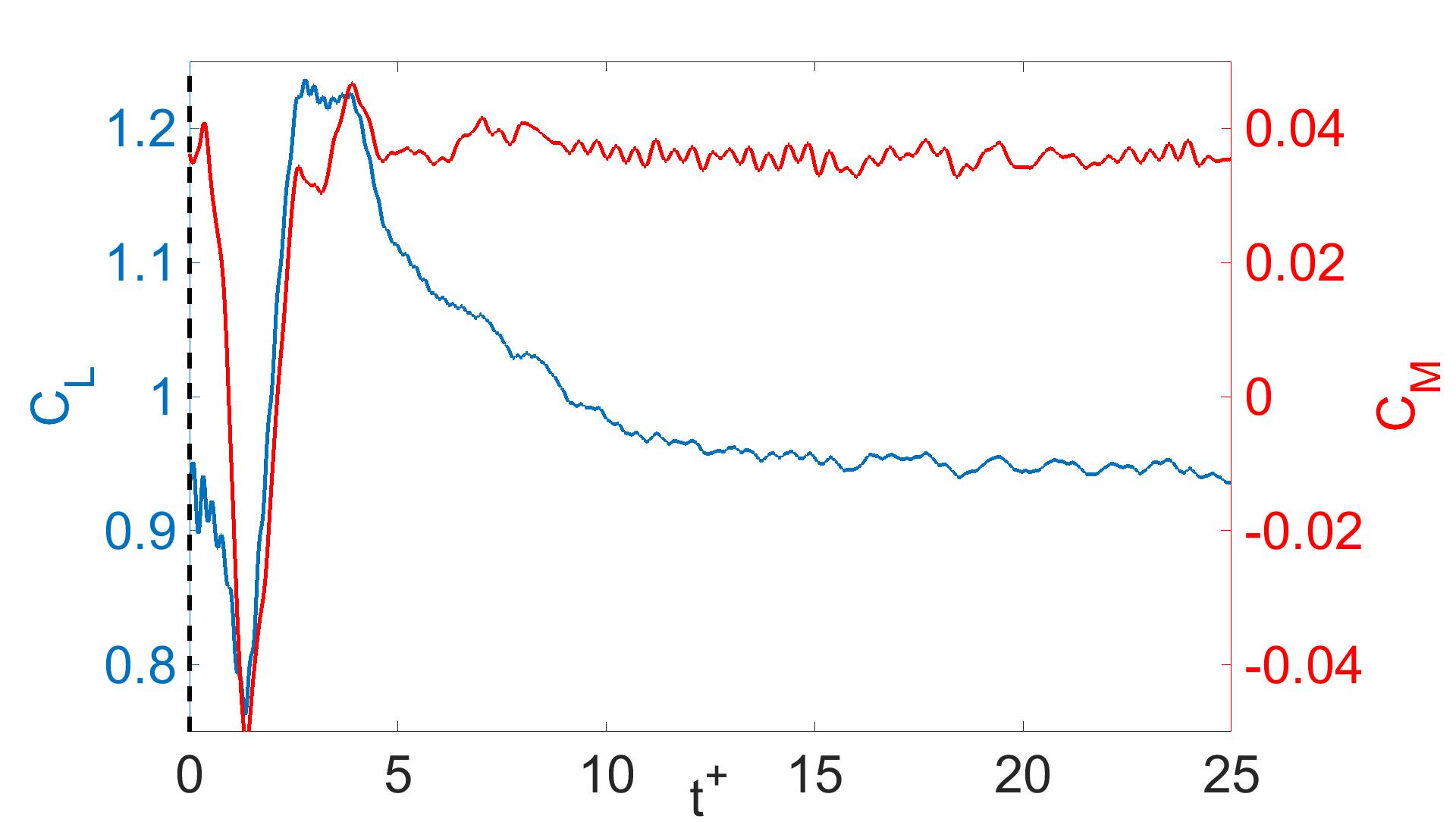}
    			\caption{$0t^+$}
    	          \label{fig:single_CL_CM_0t+}
    	\end{subfigure}	
    	~
    	\begin{subfigure}{0.3\textwidth}
    	    	  \includegraphics[width=1.7in]{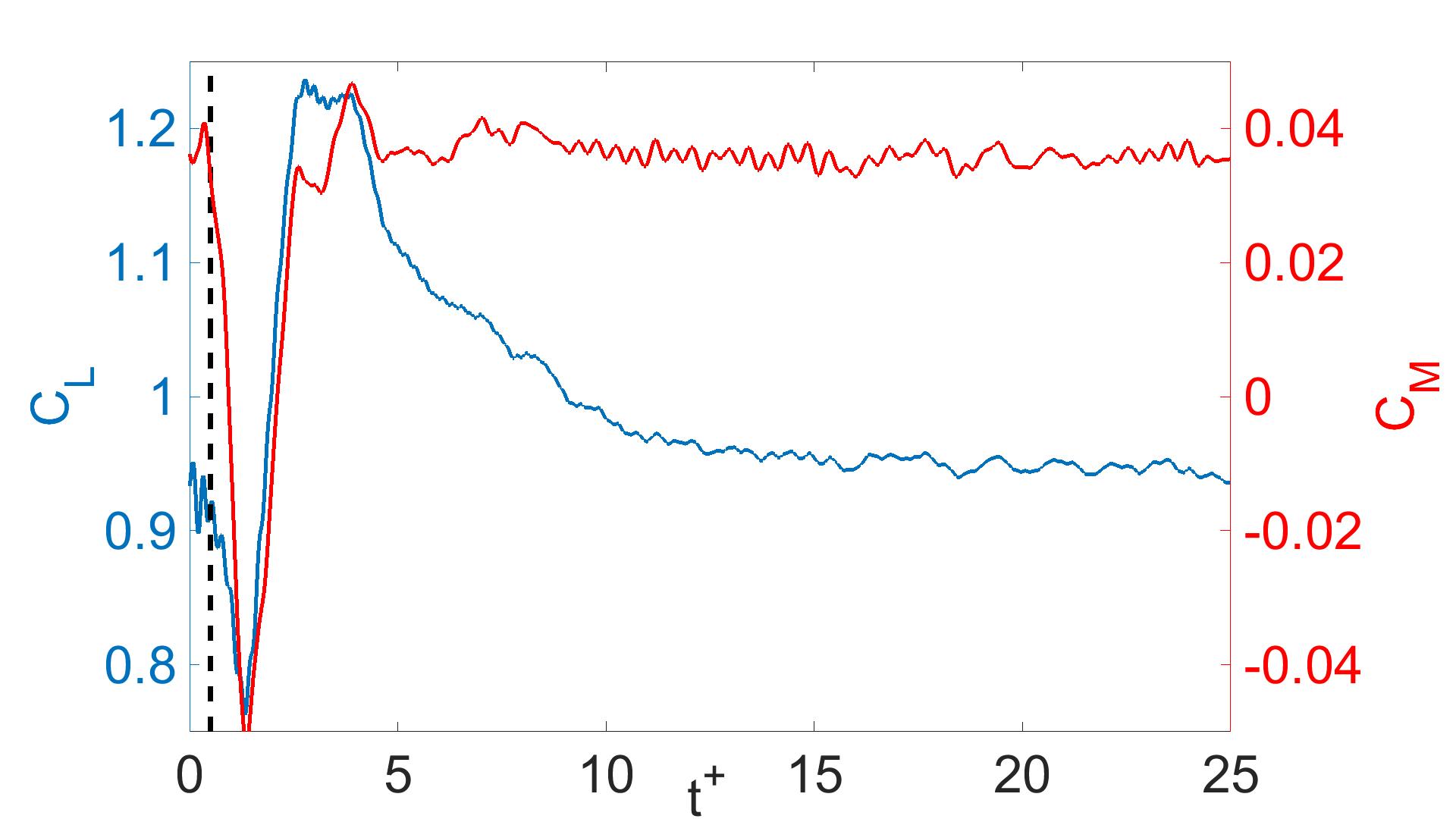}
    	    	  \caption{$0.5t^+$}
    	    	   \label{fig:single_CL_CM_0.5t+}
    	\end{subfigure}
    	~
    	\begin{subfigure}{0.3\textwidth}
    	        \includegraphics[width=1.7in]{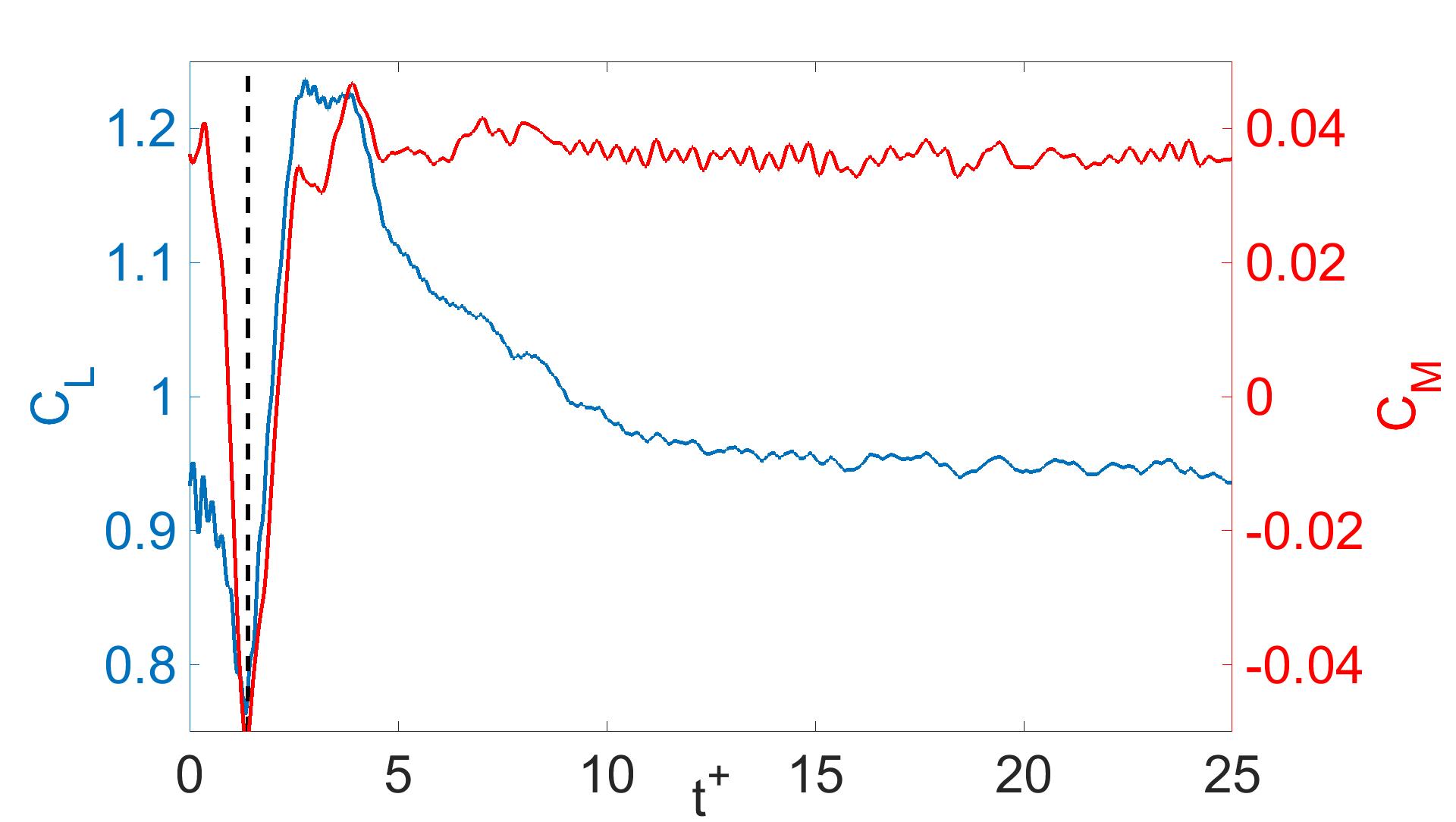}
    	          \caption{$1.4t^+$}
    	          \label{fig:single_CL_CM_1.4t+}
    	\end{subfigure}    	
 
    \begin{subfigure}{0.3\textwidth}
    	        \includegraphics[width=1.7in]{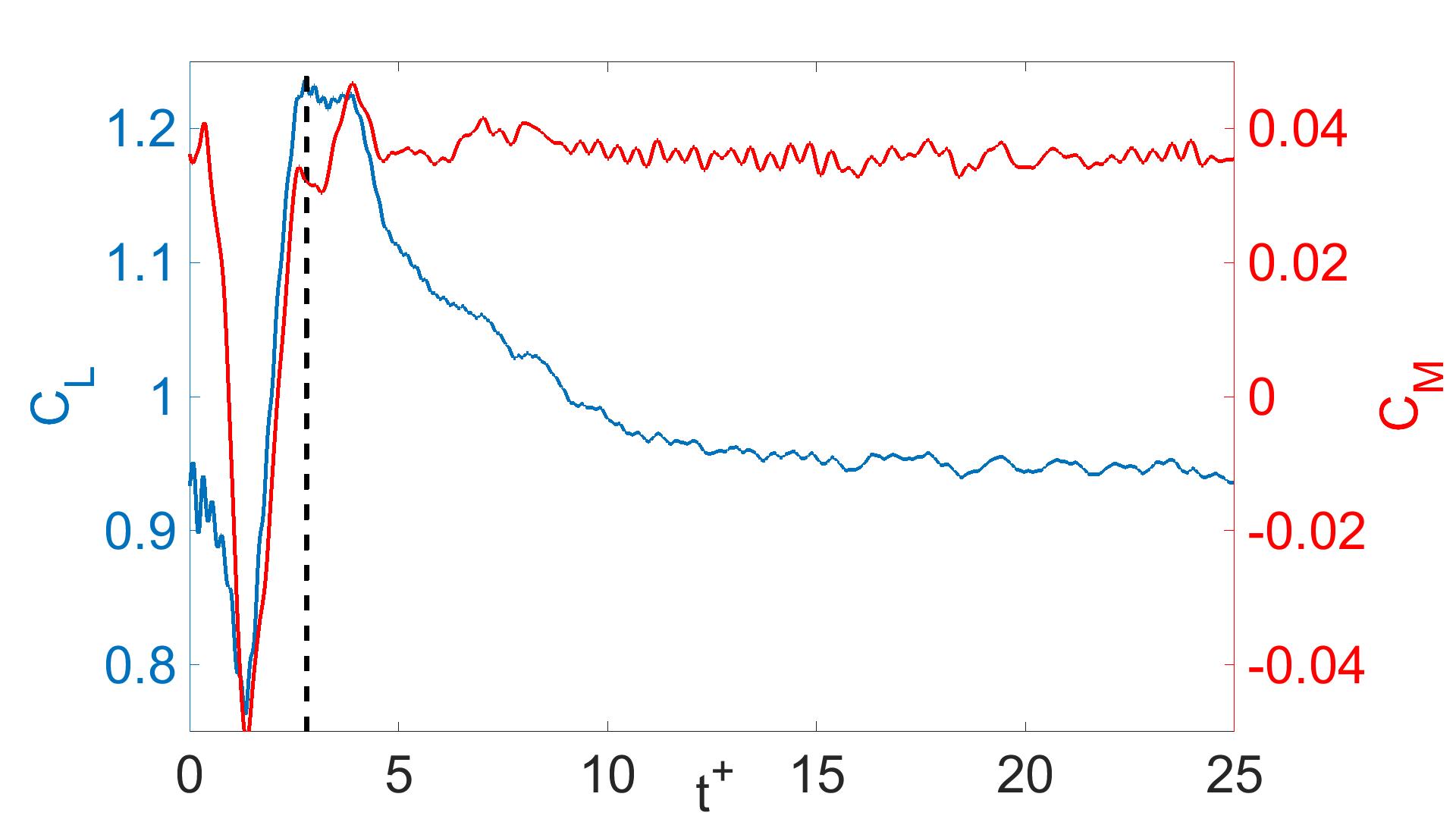}
    			\caption{$2.8t^+$}
    	          \label{fig:single_CL_CM_2.8t+}
    	\end{subfigure}	
    	~
    	\begin{subfigure}{0.3\textwidth}
    	    	  \includegraphics[width=1.7in]{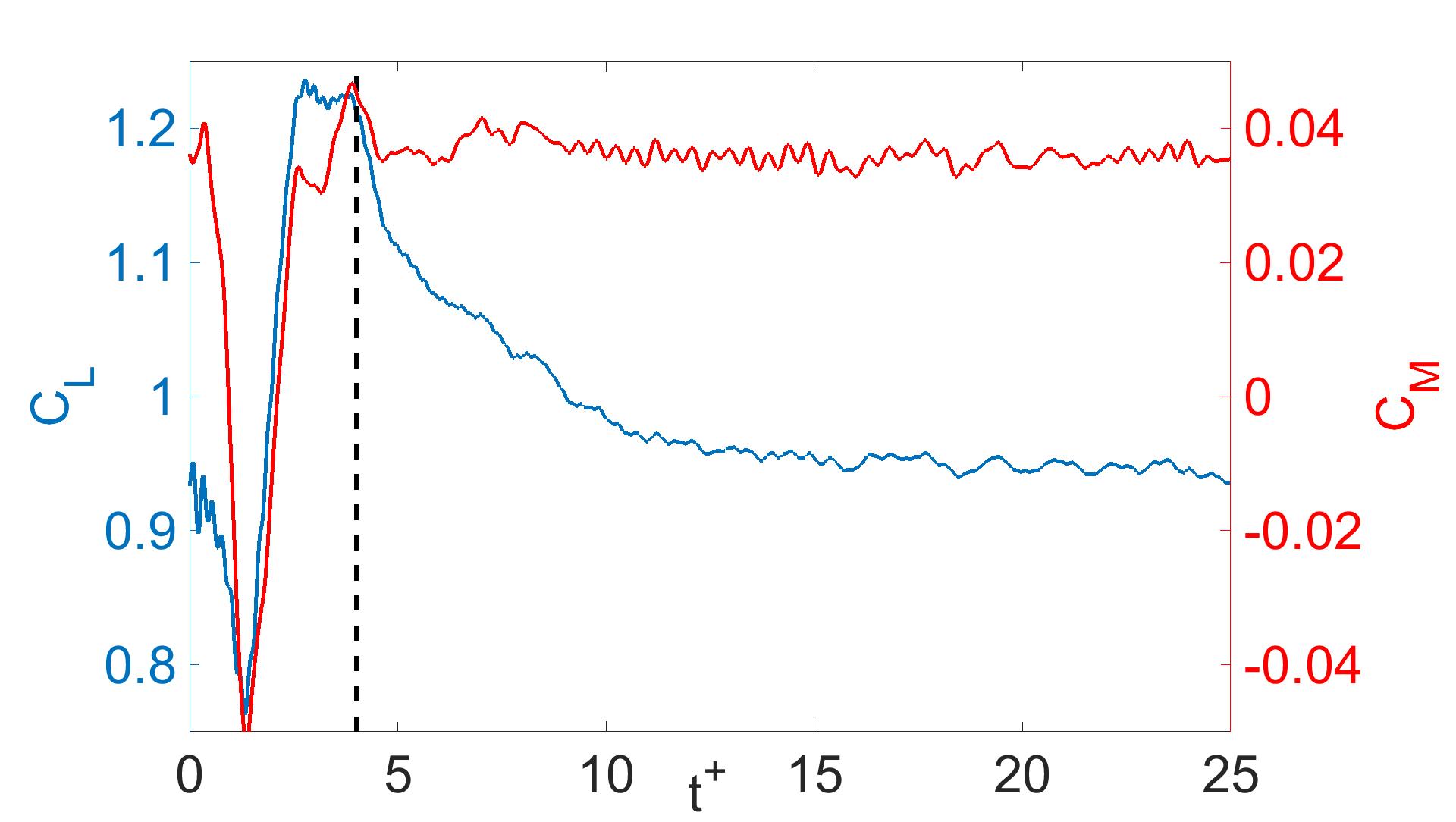}
    	    	  \caption{$4t^+$}
    	    	   \label{fig:single_CL_CM_4t+}
    	\end{subfigure}
    	~
    	\begin{subfigure}{0.3\textwidth}
    	        \includegraphics[width=1.7in]{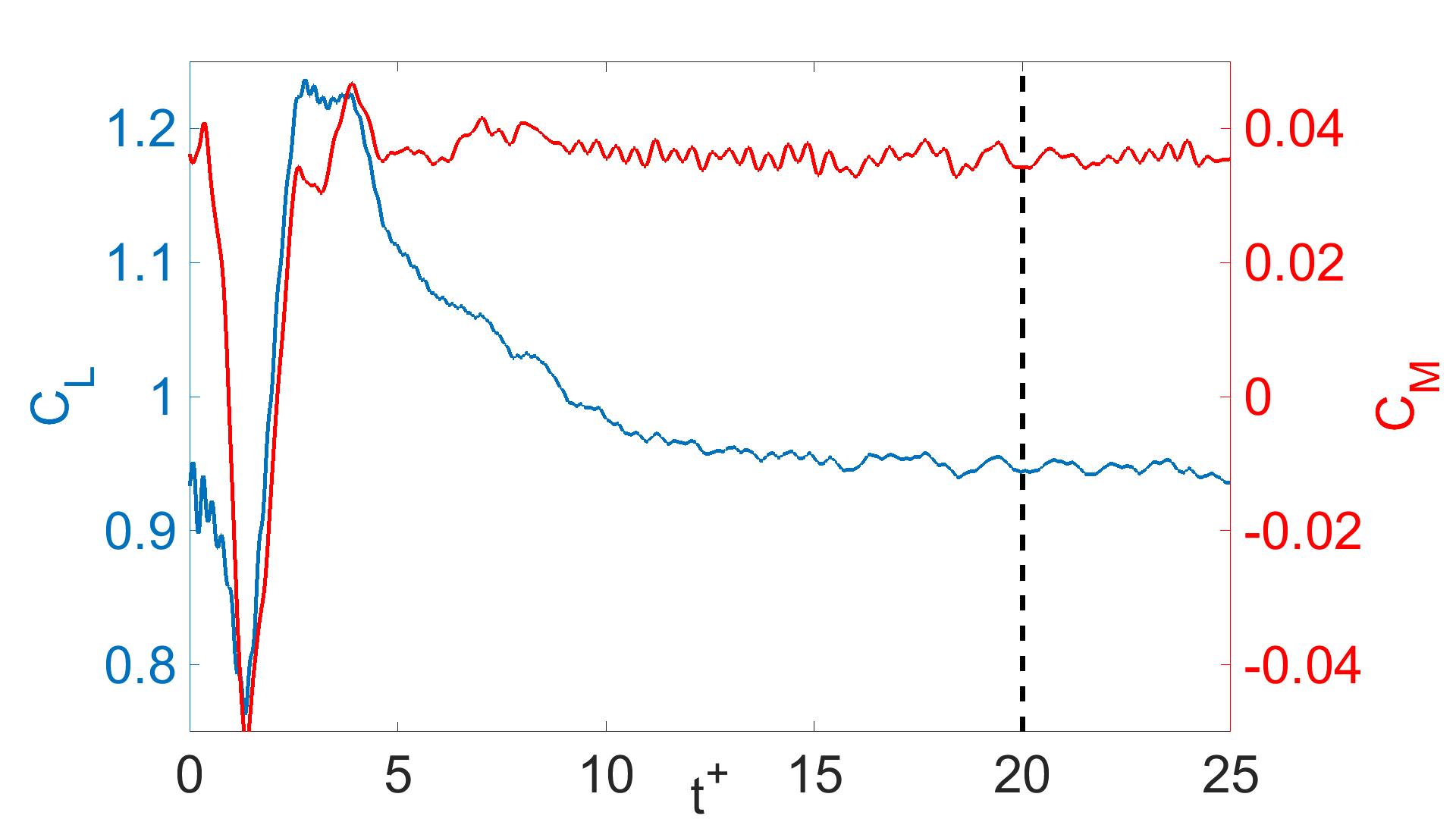}
    	          \caption{$20t^+$}
    	          \label{fig:single_CL_CM_20t+}
    	\end{subfigure}

    \caption{Phase-averaged $C_L$ (blue) and $C_M$ (red) variation following a single-burst actuation that is initiated at $0t^+$. The solid blue line denotes $C_L$, the solid red line denotes $C_M$ and the dashed black line indicates the time instant correlated to each flowfield snapshot in figure \ref{fig:flow_field}.} 
    \label{fig:single_CL_CM} 
\end{figure}

The single-burst actuation consists of a sequence of high-frequency pulses as shown in figure \ref{fig:input_single} and it is initialized at $0t^+$. The corresponding lift coefficient $C_L$ and pitching moment coefficient $C_M$ responses are shown in figure \ref{fig:single_CL_CM}. Note that we define positive $C_M$ corresponds to a nose-up pitching moment for better presentation, although it contradicts the construction of the coordinates. The vertical dashed black lines in figure \ref{fig:single_CL_CM} indicate the critical instants on the time axes, which will be discussed in more detail later. Figure \ref{fig:single_CL_CM} exhibits the lift and pitching moment reversal, or in other words, both $C_L$ and $C_M$ decrease immediately following the single-burst actuation. A similar lift reversal phenomenon was identified first by \citet{amitay2002controlled}. Since then the effect has been observed by numerous investigators, \citep{GW, brzozowski2010transient, woo2009transitory}, and is now an established feature of the separated flow dynamics. \citet{williams2018alleviating} further concluded that the lift reversal is an inherent characteristic regardless of the types of actuators or actuation.  

Figure \ref{fig:single_CL_CM_1.4t+} shows that both $C_L$ and $C_M$ reach their minima at $1.4t^+$ during the lift and pitching moment reversal. Following the lift reversal, the $C_L$ reaches its maximum at $2.8t^+$ (figure \ref{fig:single_CL_CM_2.8t+}), the maximum $C_L$ increment is about 30\% of its undisturbed baseline value. In contrast to the $C_L$ variation, there is no significant increase in $C_M$ above the baseline during the entire process (figure \ref{fig:single_CL_CM}). After $4t^+$, as it can be seen in figure \ref{fig:single_CL_CM_4t+} and figure \ref{fig:single_CL_CM_20t+} both $C_L$ and $C_M$ start to return to their baseline undisturbed condition. However, it takes longer for $C_L$ to relax compared with $C_M$.

\FloatBarrier

\begin{figure}
	\centering
    \begin{subfigure}{0.3\textwidth}
    	        \includegraphics[width=2in]{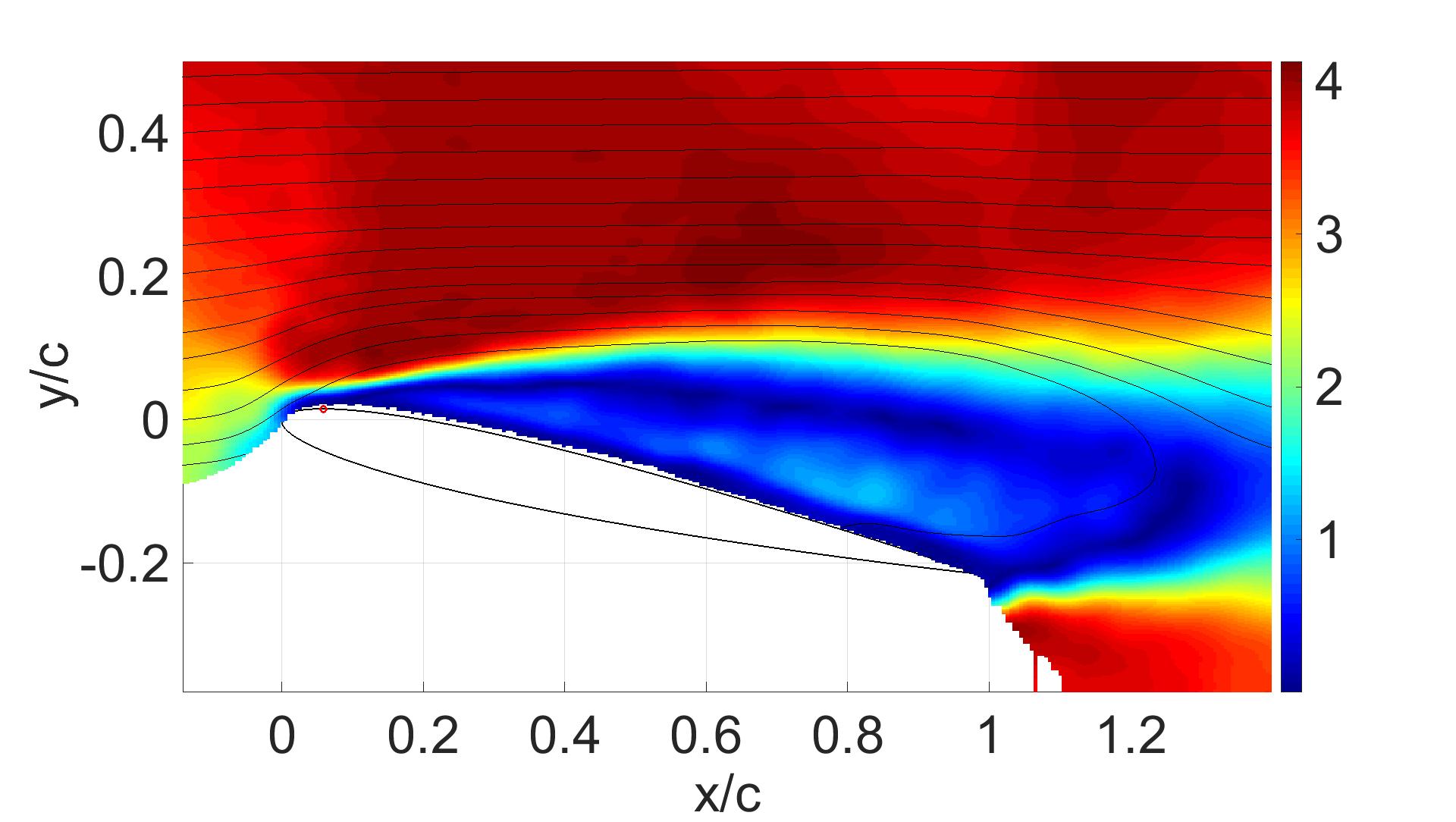}
    			\caption{$0t^+$}
    	          \label{fig:flow_field_0t+}
    	\end{subfigure}	
    	~
    	\begin{subfigure}{0.3\textwidth}
    	    	  \includegraphics[width=2in]{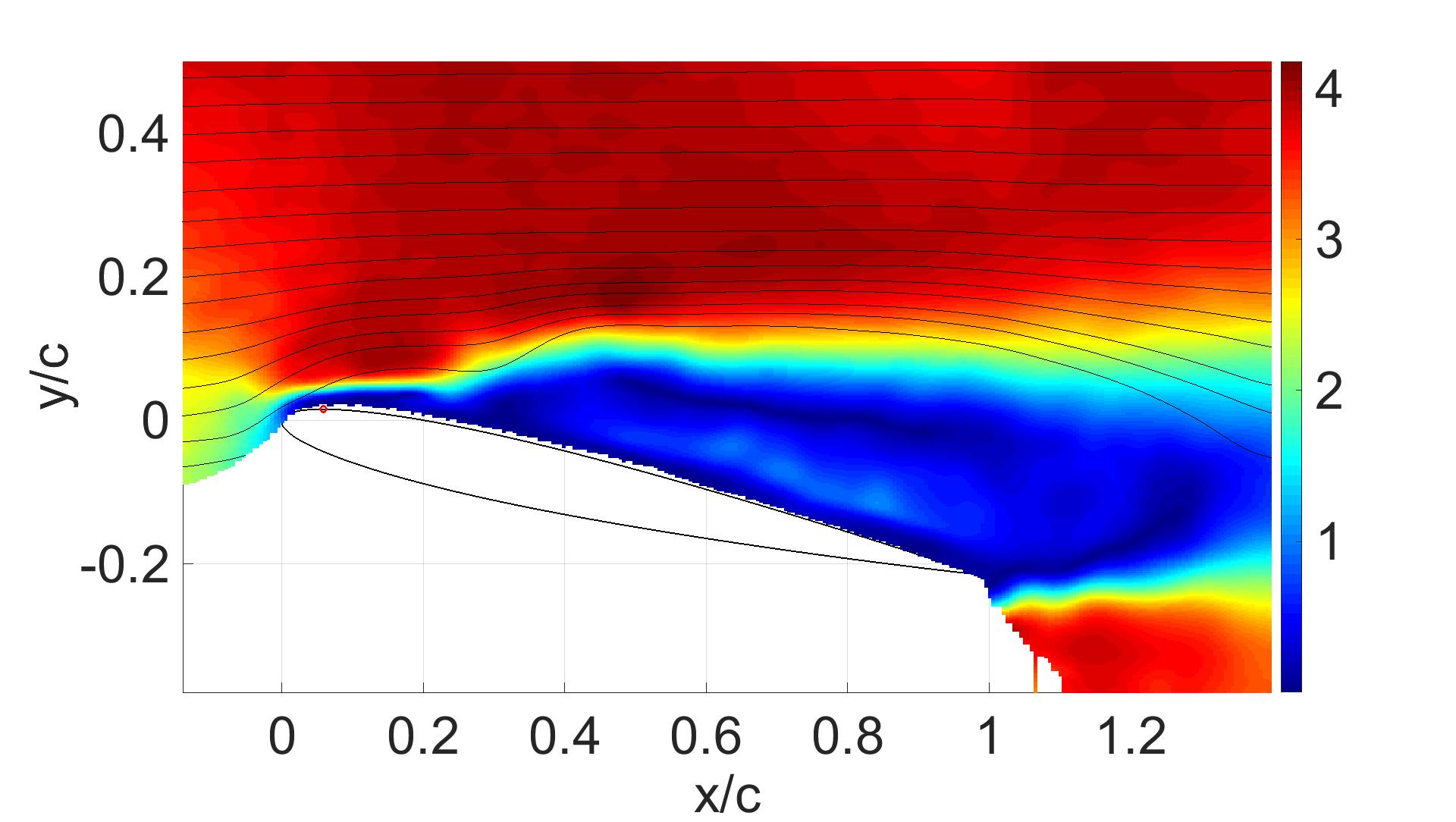}
    	    	  \caption{$0.5t^+$}
    	    	   \label{fig:flow_field_0.5t+}
    	\end{subfigure}
    	~
    	\begin{subfigure}{0.3\textwidth}
    	        \includegraphics[width=2in]{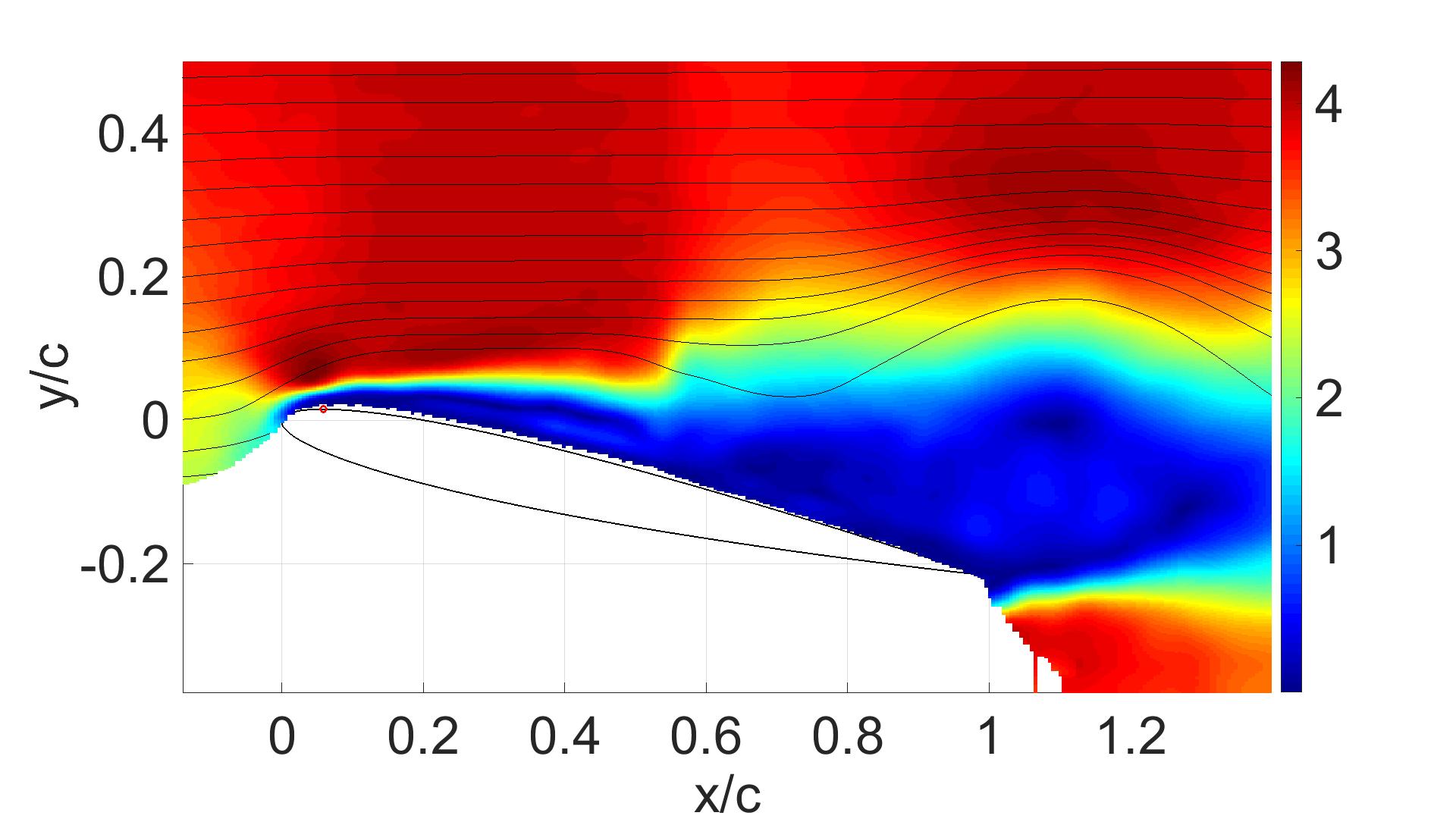}
    	          \caption{$1.4t^+$}
    	          \label{fig:flow_field_1.4t+}
    	\end{subfigure}    	
 
    \begin{subfigure}{0.3\textwidth}
    	        \includegraphics[width=2in]{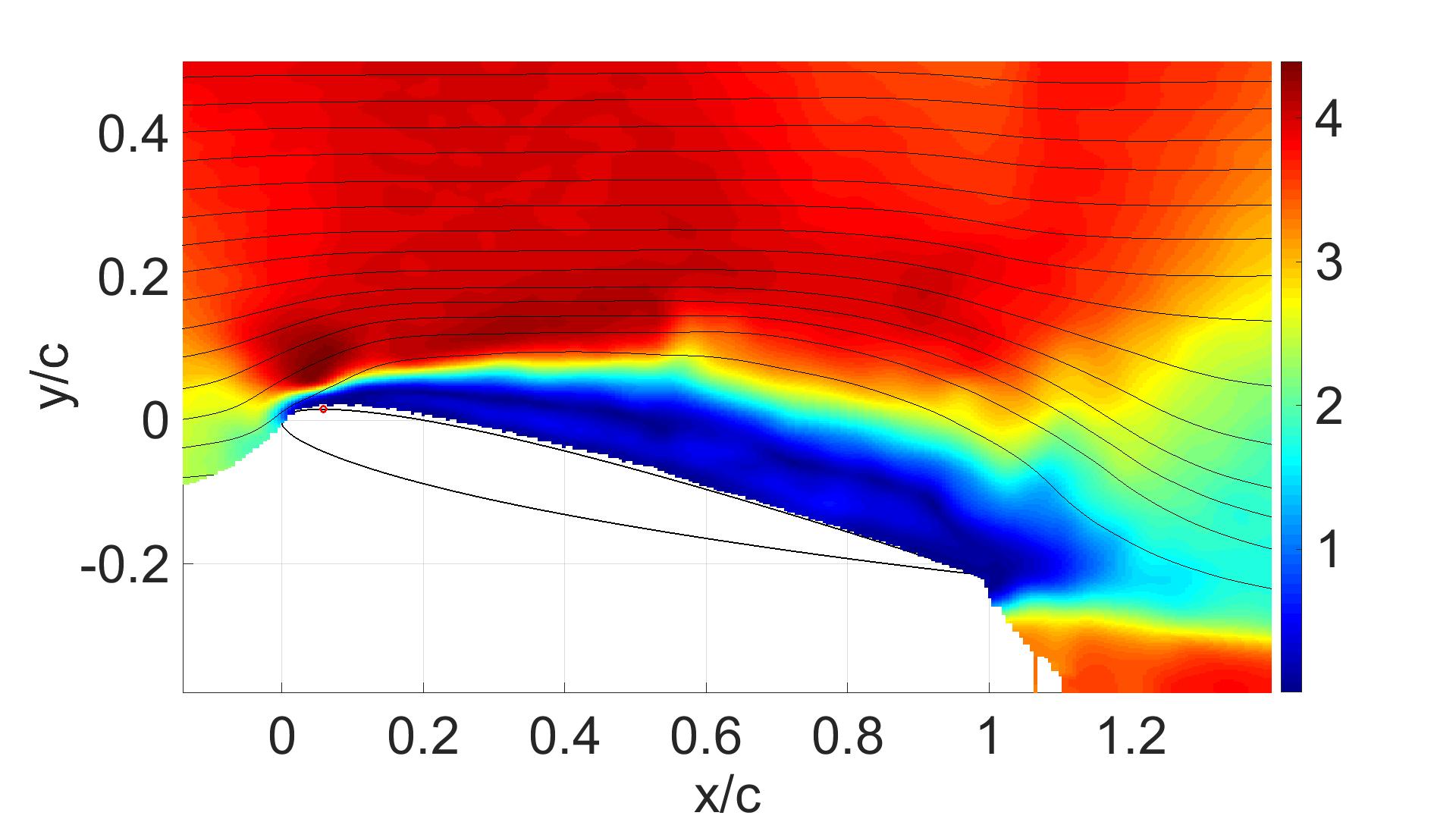}
    			\caption{$2.8t^+$}
    	          \label{fig:flow_field_2.8t+}
    	\end{subfigure}	
    	~
    	\begin{subfigure}{0.3\textwidth}
    	    	  \includegraphics[width=2in]{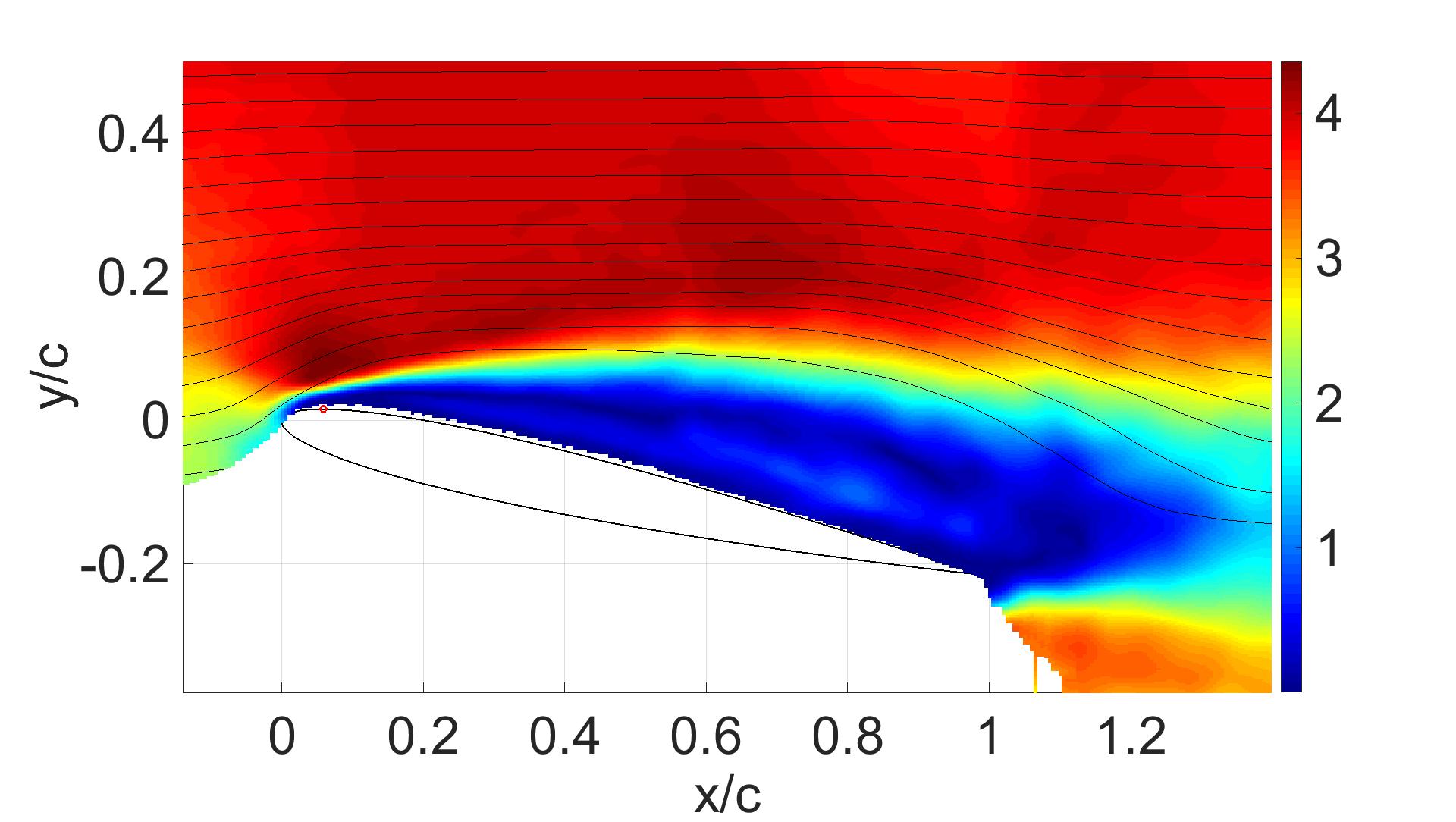}
    	    	  \caption{$4t^+$}
    	    	   \label{fig:flow_field_4t+}
    	\end{subfigure}
    	~
    	\begin{subfigure}{0.3\textwidth}
    	        \includegraphics[width=2in]{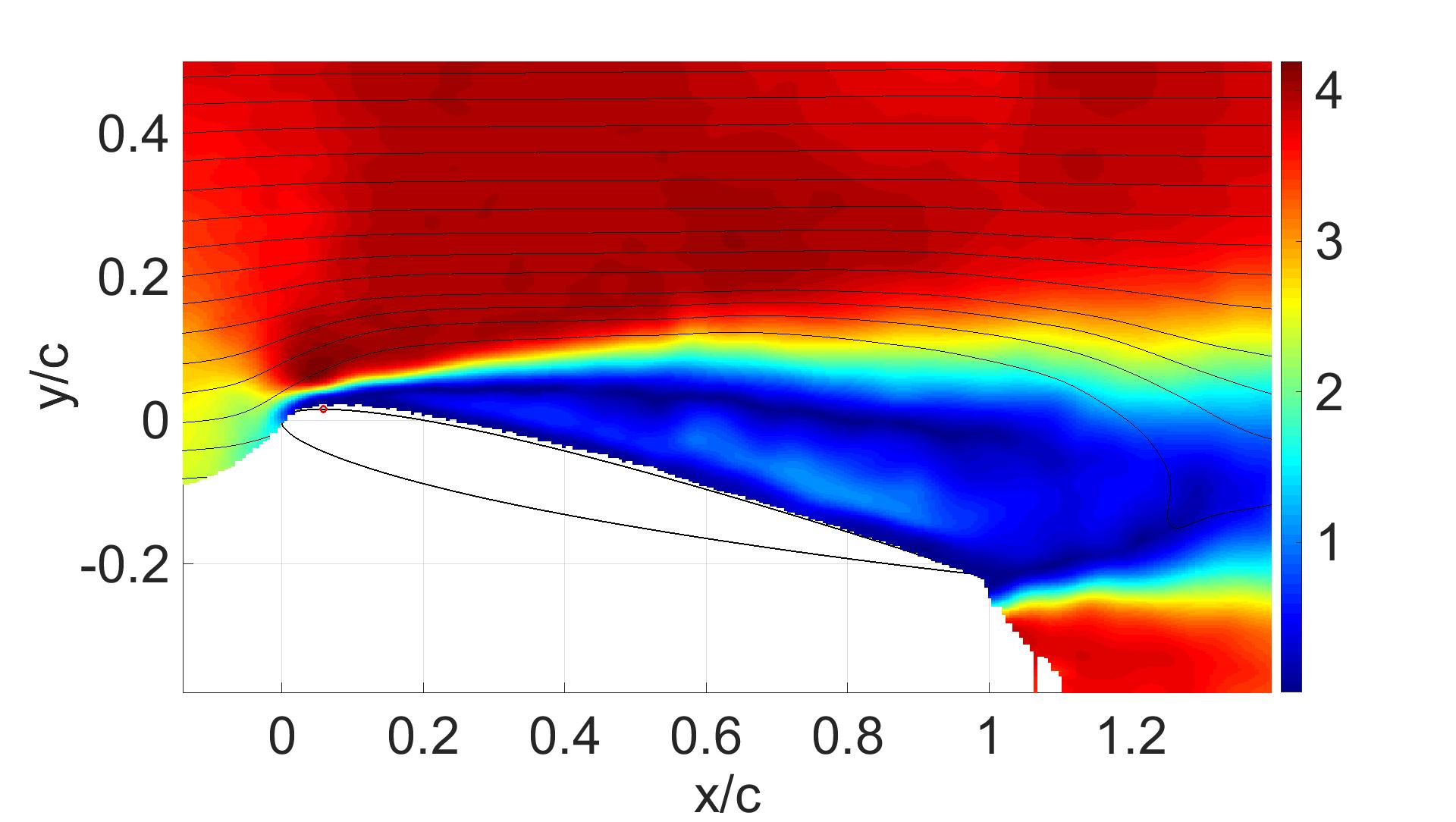}
    	          \caption{$20t^+$}
    	          \label{fig:flow_field_20t+}
    	\end{subfigure}

    \caption{Phase-averaged velocity magnitude $(\sqrt{U^2+V^2})$ and streamline time sequence plots following a single-burst actuation that is initiated at $0t^+$, where $U$ is the horizontal velocity along $x$ and $V$ is the vertical velocity along $y$. The color indicates the velocity magnitude and the black lines are showing the streamline. The red circle on the leading edge denotes the streamwise location of the actuators. The flow direction is from left to right.} 
    \label{fig:flow_field} 
\end{figure}

The corresponding velocity field and streamline at the critical time instants following the single-burst actuation are shown in figure \ref{fig:flow_field} with respect to the time instants marked by the dashed black lines in figure \ref{fig:single_CL_CM}. In this figure, the color corresponds to velocity magnitude. Referring to the velocity field prior to the burst, the baseline flow on the suction side is fully separated at $\alpha=12^o$ angle of attack (figure \ref{fig:flow_field_0t+}). The single-burst actuation was initiated at $0t^+$ and lasted for $0.12t^+$. The beginning of the reattachment process can be seen at  $0.5t^+$ at $x/c=0.2$ from the leading edge (figure \ref{fig:flow_field_0.5t+}). The reattachment produces a “kink” in the shear layer that divides the new `reattached' flow from the `old' separated flow region. The reattached region grows with time as the kink convects downstream from the leading edge towards the trailing edge as shown in figure \ref{fig:flow_field_0.5t+} to figure \ref{fig:flow_field_2.8t+}.  The maximum flow reattachment occurs when the kinked region of the shear layer reaches the trailing edge at $2.8t^+$ (figure \ref{fig:flow_field_2.8t+}). At this time, the lift coefficient also reaches its maximum (figure \ref{fig:single_CL_CM_2.8t+}), which agrees with Rival's observations \citep{rival2014characteristic}.  After $4t^+$, the lift coefficient begins to decrease as the flowfield gradually relaxes to its original baseline state. This relaxation process is exhibited in figure \ref{fig:flow_field_4t+} and figure \ref{fig:flow_field_20t+}, and it takes about $10t^+$. \citet{brzozowski2010transient} described much of the same behavior when using combustion-based pulsed actuators on a NACA 4415 cambered wing.
\FloatBarrier

\section{The mechanism of $C_L$, $C_M$ reversal}\label{sec:reversal}

The mechanism behind the lift and pitching moment reversals is important to be understood because this is the cause of the non-minimum phase behavior that limits control bandwidth.

\subsection{The relation between $C_L$,$C_M$, Pressure and the flowfield}

 By examining the vortex structure and surface pressure evolution, we intend to gain some insight into the physics of the lift and pitching moment reversal. The methodology here is that we will first analyze the relation between lift, pitching moment and the pressure, and then analyze the relation between the surface pressure and the flowfield. 
 
\begin{figure}

\centering
	\begin{subfigure}{0.4\textwidth}
    	\centering
    	    	  \includegraphics[width=1\textwidth]{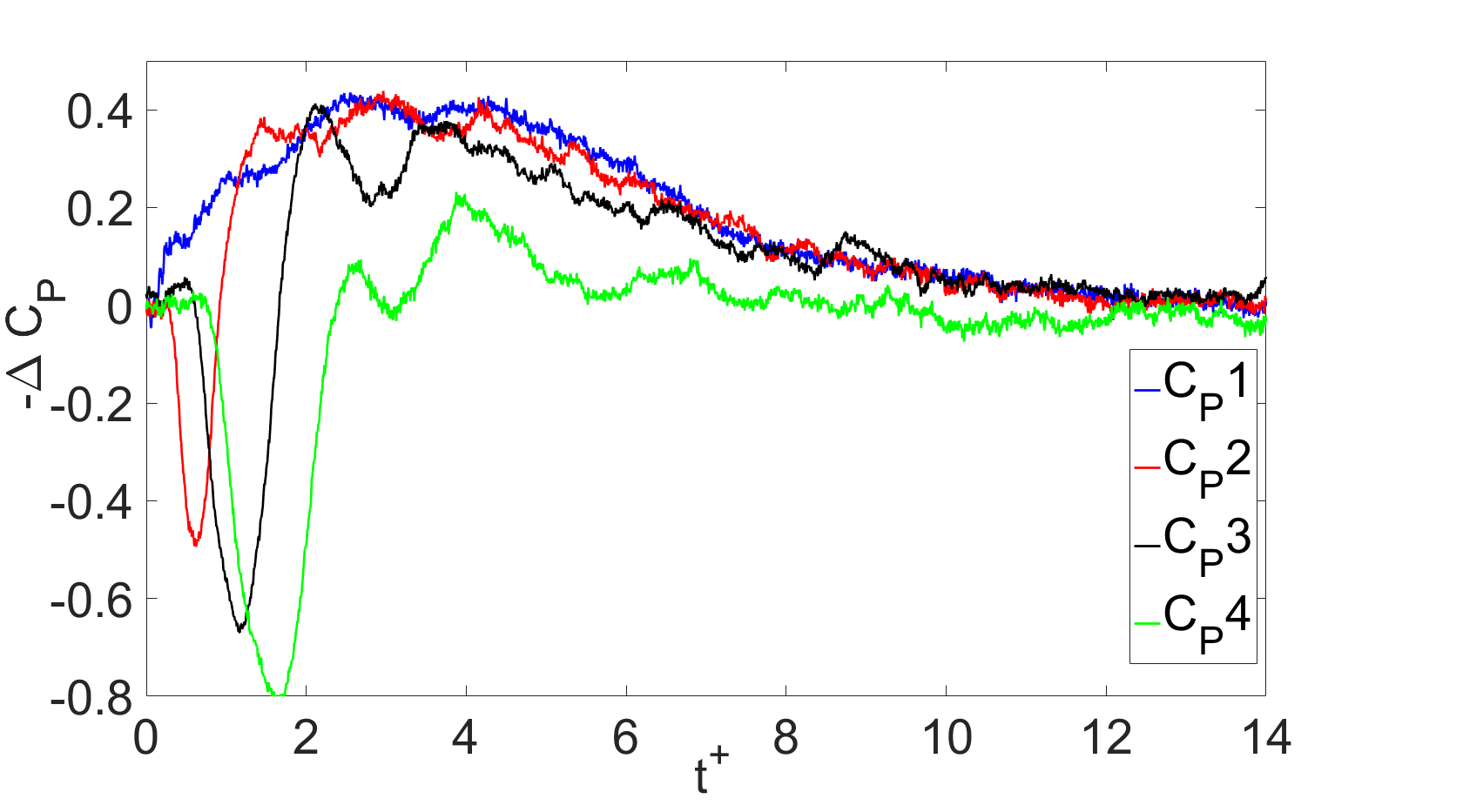}
    	    	  \caption{$\Delta C_P$ following the single-burst actuation.}
    	    	   \label{fig:single_pressure}
    	\end{subfigure}
    	~
    \begin{subfigure}{0.4\textwidth}
    \centering
    
    	\includegraphics[width=1\textwidth]{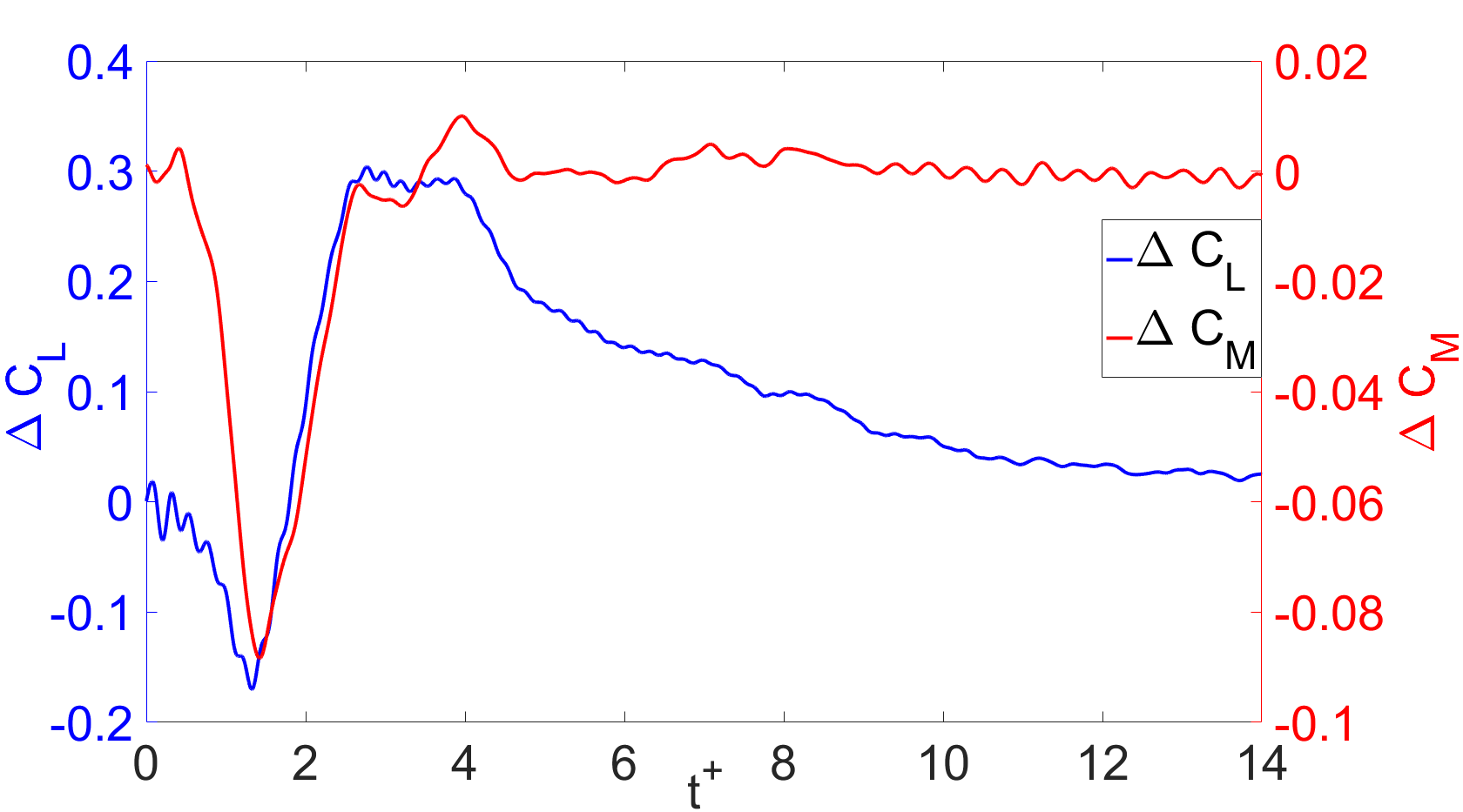}
    			\caption{$\Delta C_L$, and $\Delta C_M$ following the single-burst actuation.}
    	          \label{fig:single_DCL_DCM}
    	\end{subfigure}

\caption{Lift coefficient increment and pitching moment coefficient increment measured by the force transducer, and pressure coefficient increments measured by the four pressure sensors.}
\label{fig:CL_CM_pressure}
\end{figure}

 The time series data of the increments, $\Delta C_P$, $\Delta C_L$ and $\Delta C_M$ are shown in figure \ref{fig:CL_CM_pressure}. Here, $"\Delta"$ denotes the value in the actuated cases relative to the non-actuated baseline. Note that the vertical axis on figure \ref{fig:single_pressure} is $"-\Delta C_P"$ for a easier comparison with $\Delta C_L$ and $\Delta C_M$ (figure \ref{fig:single_DCL_DCM}). The pressure response to the single-burst actuation was investigated at four chordwise locations. It can be seen in figure \ref{fig:single_pressure} that the pressure at locations PS2, PS3, and PS4 follows a similar trend after a burst disturbance from the actuator is initiated. The more important observation is that the pressure on these three pressure sensors shows the reversal-like phenomenon, which is very similar to the lift/pitching moment reversal shown in figure \ref{fig:single_DCL_DCM}, there is an approximately $0.6 t^+$ constant time delay between the minima on the pressure sensors due to the convection of the disturbance. It is also worth of noting that the first pressure signal on PS1 does not follow the same pattern as the other pressure sensors, because it is located upstream of the actuator. 

\FloatBarrier

\begin{figure}
\centering
    	\begin{subfigure}{0.45\textwidth}
    	\centering
		\includegraphics[width=1\textwidth]{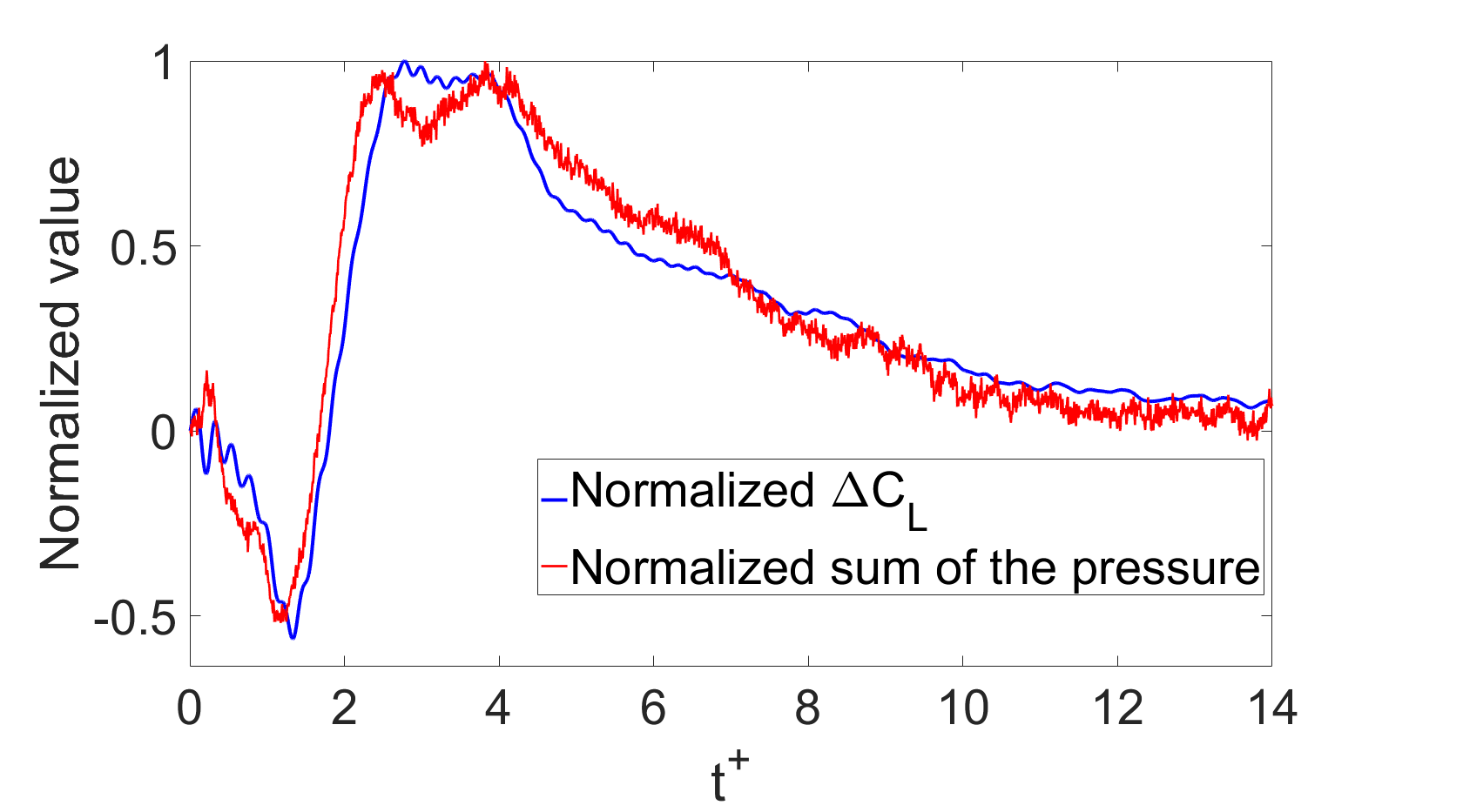}
			\caption{Comparison of $\Delta C_L$ normalized by the maximum lift increment, and the sum of the all 4 pressure measurements normalized by the maximum.}
	    \label{fig:DCL_single_pressure}
	    \end{subfigure}
~
    	\begin{subfigure}{0.45\textwidth}
    	\centering
		\includegraphics[width=1\textwidth]{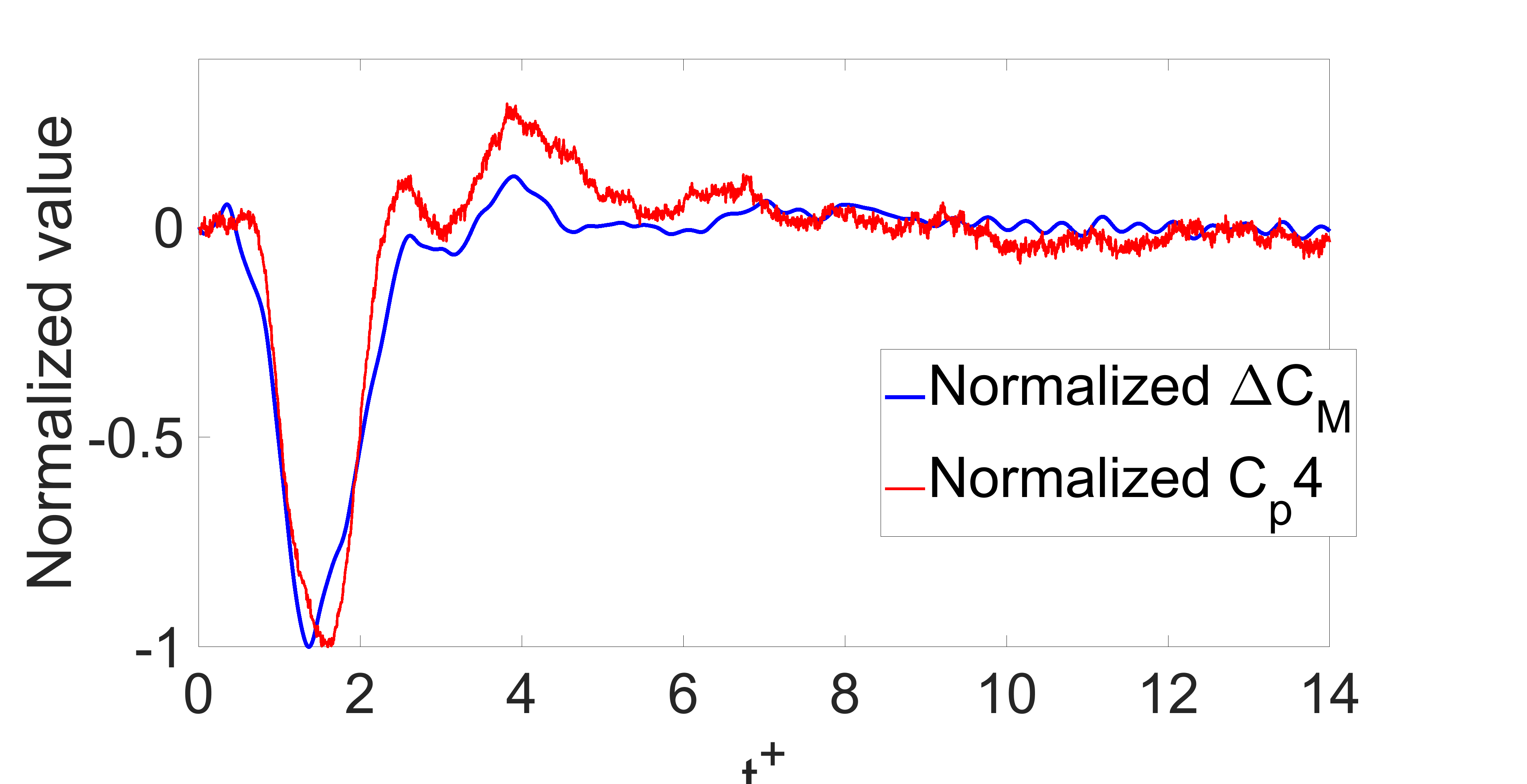}
			\caption{Comparison of the normalized $\Delta C_M$ and the normalized $C_P4$.}
	    \label{fig:DCM_single_pressure}
	    \end{subfigure}
\caption{$\Delta C_L$, $\Delta C_M$ and $\Delta C_P$ normalized by their respective maxima.}
\end{figure}

To better visualize the relation between $\Delta C_L$ and $\Delta C_P$, a comparison of $\Delta C_L$ normalized by its maximum (blue) and the sum of $\Delta C_P$ on all pressure sensors normalized by its maximum is plotted in figure \ref{fig:DCL_single_pressure} (red). It is obvious that $\Delta C_L$ closely follows the trend of the sum of all pressure measurements. On the other hand, when both $\Delta C_M$ and $\Delta C_P4$ are normalized by their minima, we found that the trend of $\Delta C_M$ is closely tracked by $C_P4$ (figure \ref{fig:DCM_single_pressure}) measured by PS4, which is the closest to the trailing edge.  This is because the moment arm of PS4 is the largest relative to the reference point (0.25c) of the moment measurement. 

Up to now, we can conclude that the $C_L$ and $C_M$ reversal is a consequence of the surface pressure reversal following the initiation of the single-burst actuation. Next, we will connect the time-varying surface pressure to the vortex structure, so that we can obtain a more comprehensive picture of how the flowfield evolution contributes to the $C_L$, $C_M$ reversal.

\FloatBarrier

\begin{figure}
\centering
\includegraphics[width=0.5
\textwidth]{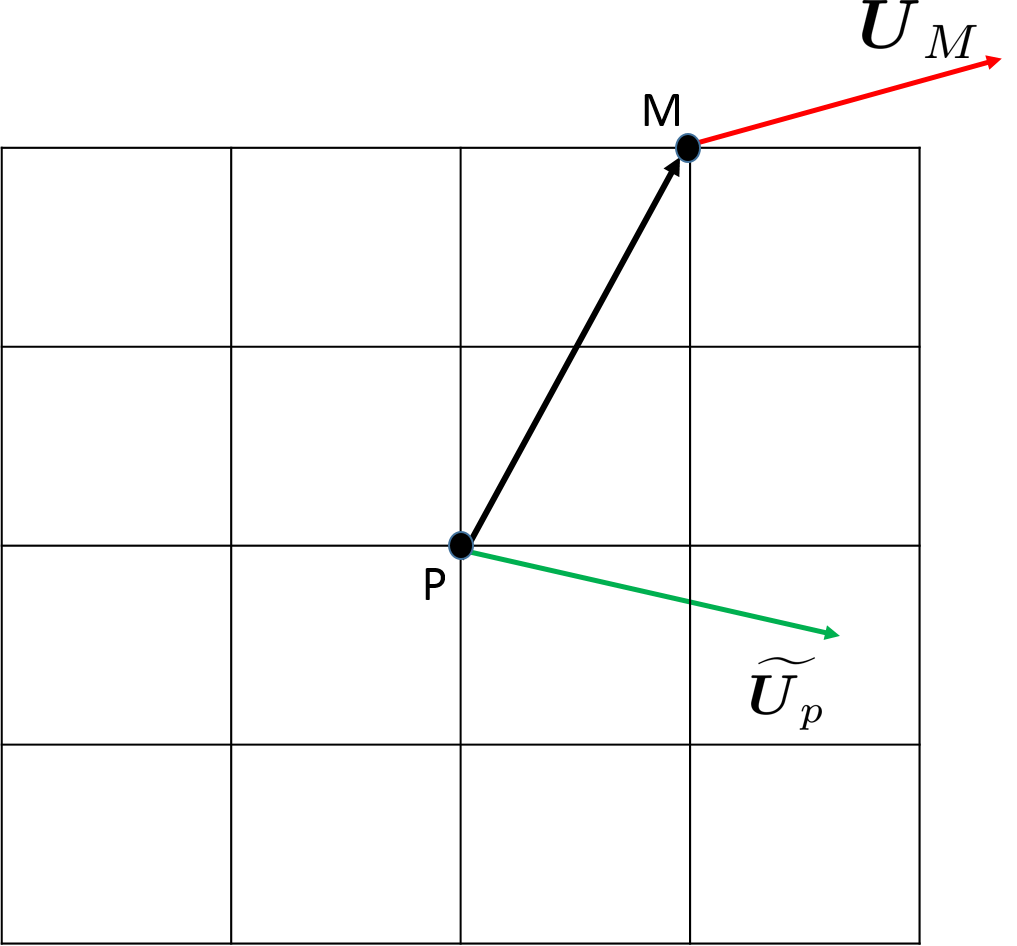}
\caption{The sketch of the $\Gamma$ computational area.}
\label{fig:Gamma2}
\end{figure}

To study the flow structure associated with the lift and pitching moment reversal, a method following \citet{graftieaux2001combining} is used to calculate the vortex strength. In \citet{graftieaux2001combining}'s work, a Galilean invariant vortex strength $\Gamma$ in the flow was calculated using the local swirling velocity and the spatial vector relative to the center point of the computational domain. To reduce the noise in the measured flow field, a local averaging method was used. The calculation of the vortex strength is shown as follow,
\begin{equation}
\label{eq:Gamma}
\Gamma(p)=\frac{1}{N}\sum_S[\boldsymbol{P}\boldsymbol{M}\wedge(\boldsymbol{U}_M-\widetilde{\boldsymbol{U}_p})]\cdot \boldsymbol{Z}
\end{equation}
where $\boldsymbol{P}\boldsymbol{M}$ is the spatial vector from the center point $P$ of the computational area to each individual point $M$ surrounding $P$ in the computational area $S$. $N$ is the number of points in the surrounding area. $\boldsymbol{U}_M$ is the velocity at the point $M$. $\widetilde{\boldsymbol{U}_p}$ is the mean velocity in the area $S$ and $\boldsymbol{Z}$ is the unit vector normal to the measurement plane. In the 2-D case, Eq. \ref{eq:Gamma} becomes
\begin{equation}
\label{eq:Gamma_2D}
\Gamma(p)=\frac{1}{N}\sum_S[\boldsymbol{P}\boldsymbol{M}\wedge(\boldsymbol{U}_M-\widetilde{\boldsymbol{U}_p})]
\end{equation}
The computational domain of $\Gamma$ at the spatial point $P$ is sketched in figure \ref{fig:Gamma2}.  
\FloatBarrier

To better visualize the vortex structure, the Galilean invariant $\Lambda_2$ criterion \citep{jeong1995identification} was used to identify the vortex boundary. The vortex strength $\Gamma$ is only shown inside the vortex boundary identified by $\Lambda_2$. A brief description of $\Lambda$ calculation is given here. Taking the gradient of Navier-Stokes equations, the symmetric part without unsteady and viscous effects is

\begin{equation}
\label{eq:lambda}
\Omega^2+S^2=-\frac{1}{\rho}\bigtriangledown(\bigtriangledown P)
\end{equation}
where $S=(J+J^T)/2$, $\Omega=(J-J^T)/2$ and $J$ is the velocity gradient tensor. The tensor $\Omega^2+S^2$ is the
corrected Hessian of pressure. The vortex core can be defined as a connected region with two negative eigenvalues $(\Lambda_2<0)$ of tensor $\Omega^2+S^2$. In our case, $(\Lambda_2<-100)$ is used to reduce the number of the small vortices that are caused by the turbulence and measurement noise.

\begin{figure}
\centering
    \includegraphics[width=0.2\textwidth]{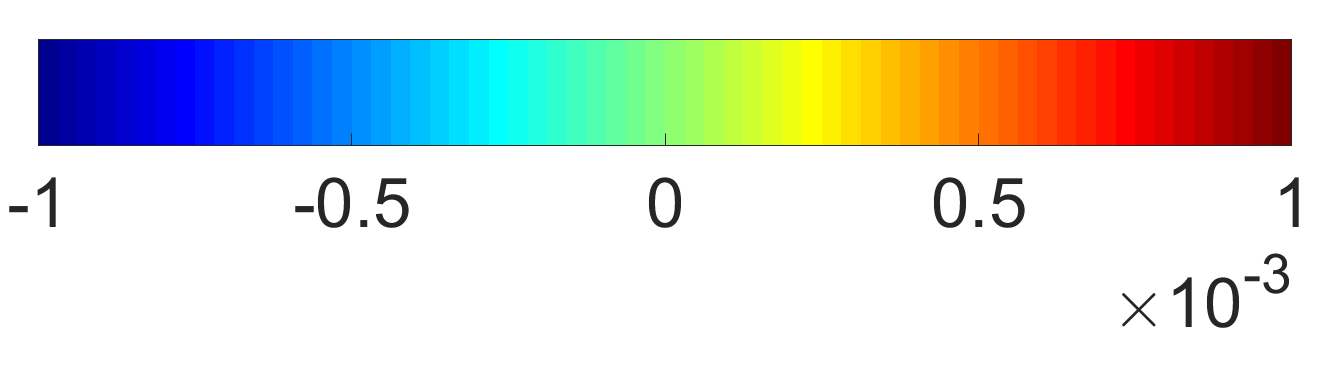}
\begin{multicols}{2}
	\centering
	\begin{subfigure}{0.45\textwidth}
	        \includegraphics[width=2.8in]{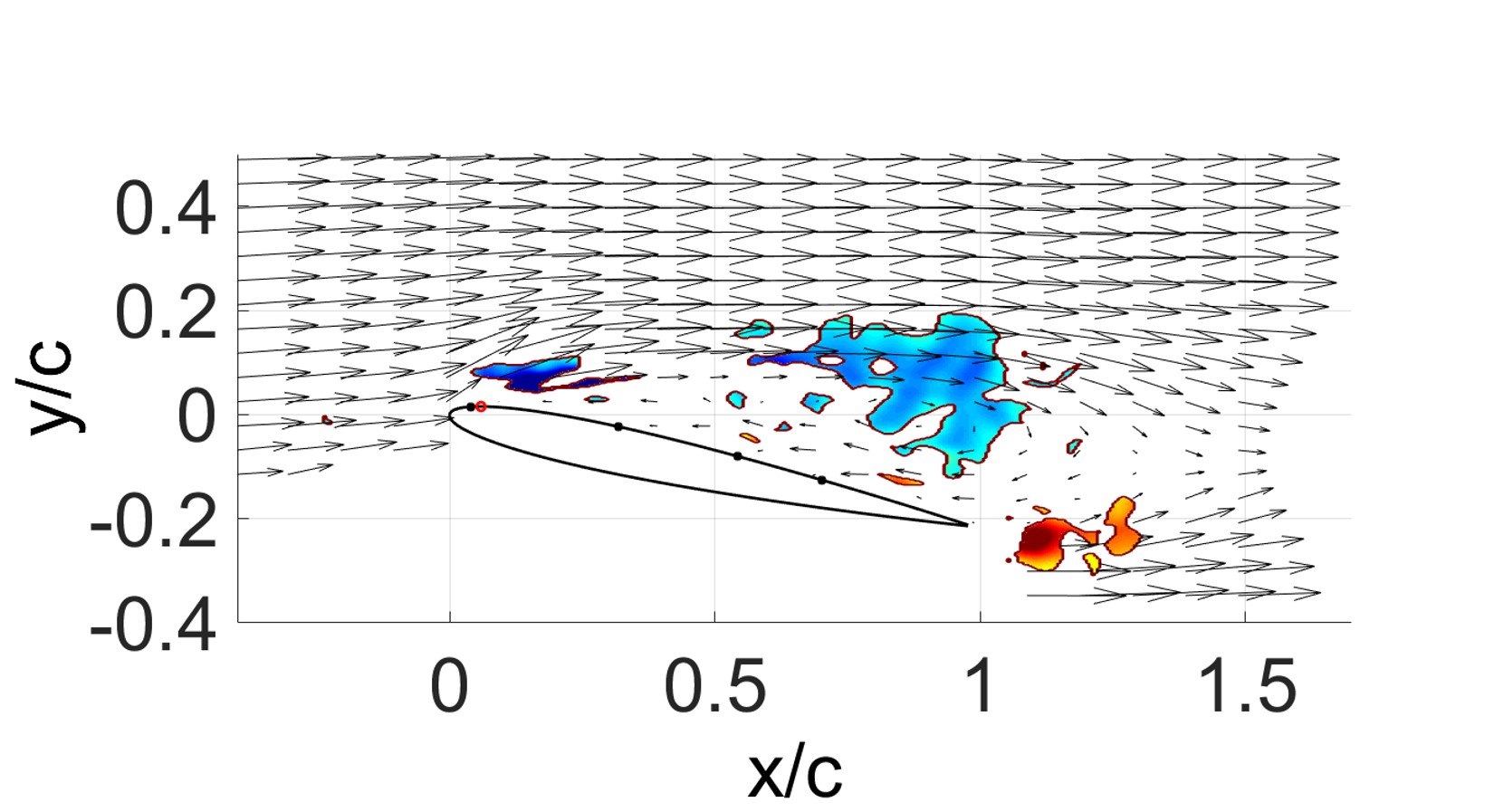}
			\caption{$0t^+$}
	          \label{fig:Gamma2_0}
	\end{subfigure}	
	
	\begin{subfigure}{0.45\textwidth}
	        \includegraphics[width=2.8in]{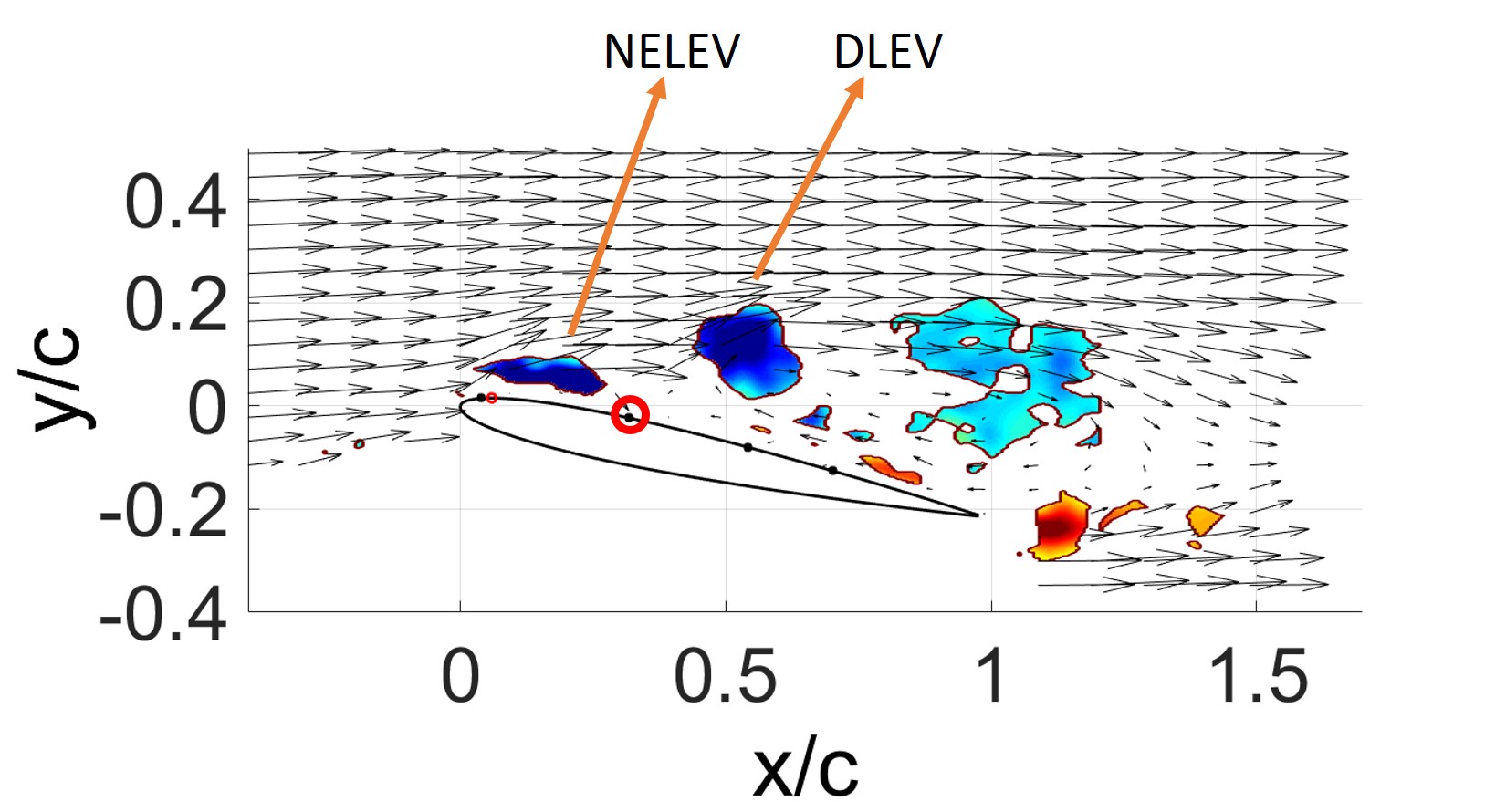}
	          \caption{$0.7t^+$}
	          \label{fig:Gamma2_0p7}
	\end{subfigure}	
	
	\begin{subfigure}{0.45\textwidth}
	        \includegraphics[width=2.8in]{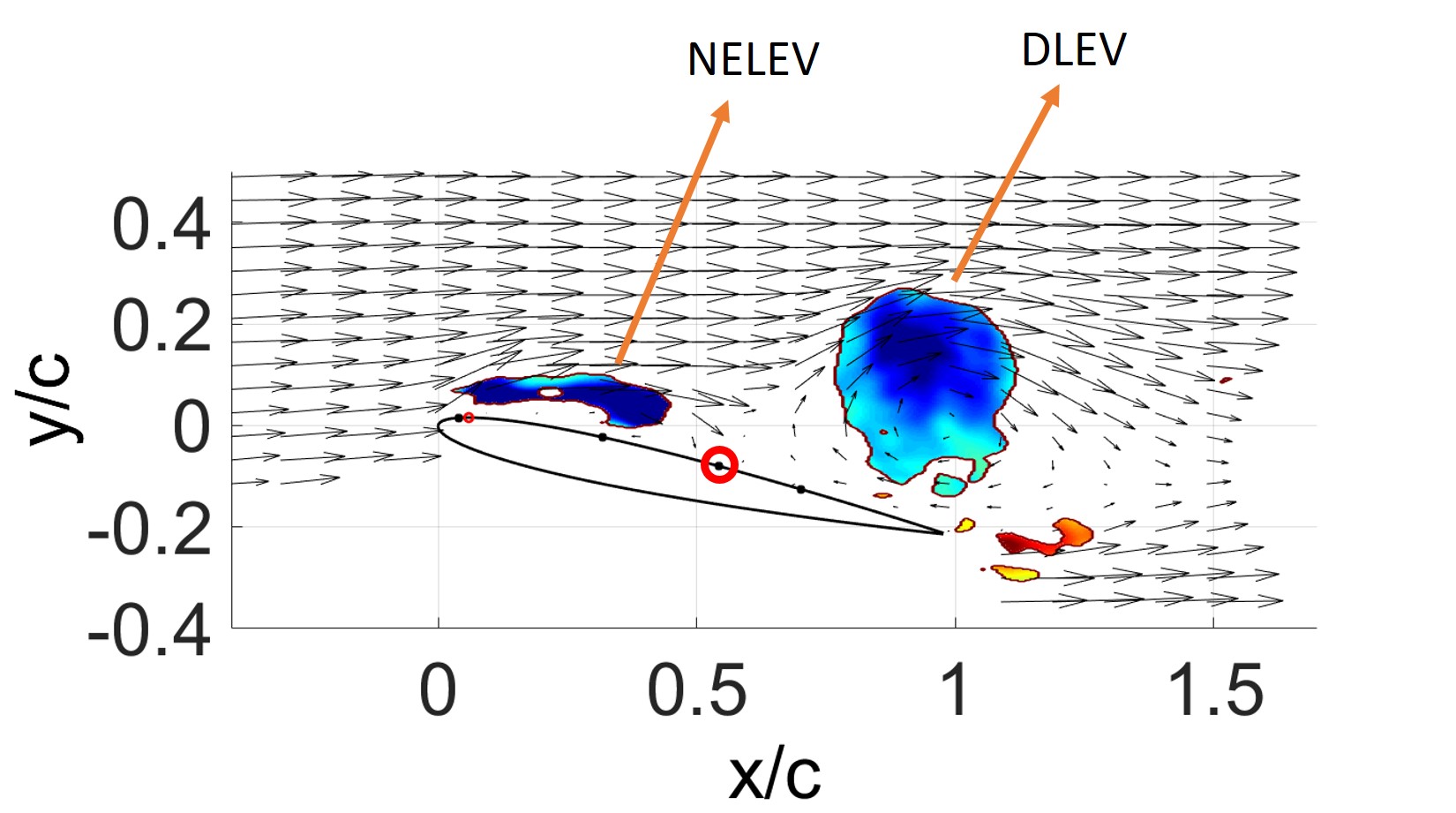}
	          \caption{$1.2t^+$}
	          \label{fig:Gamma2_1p2}
	\end{subfigure}	
	
	\begin{subfigure}{0.45\textwidth}
	        \includegraphics[width=2.8in]{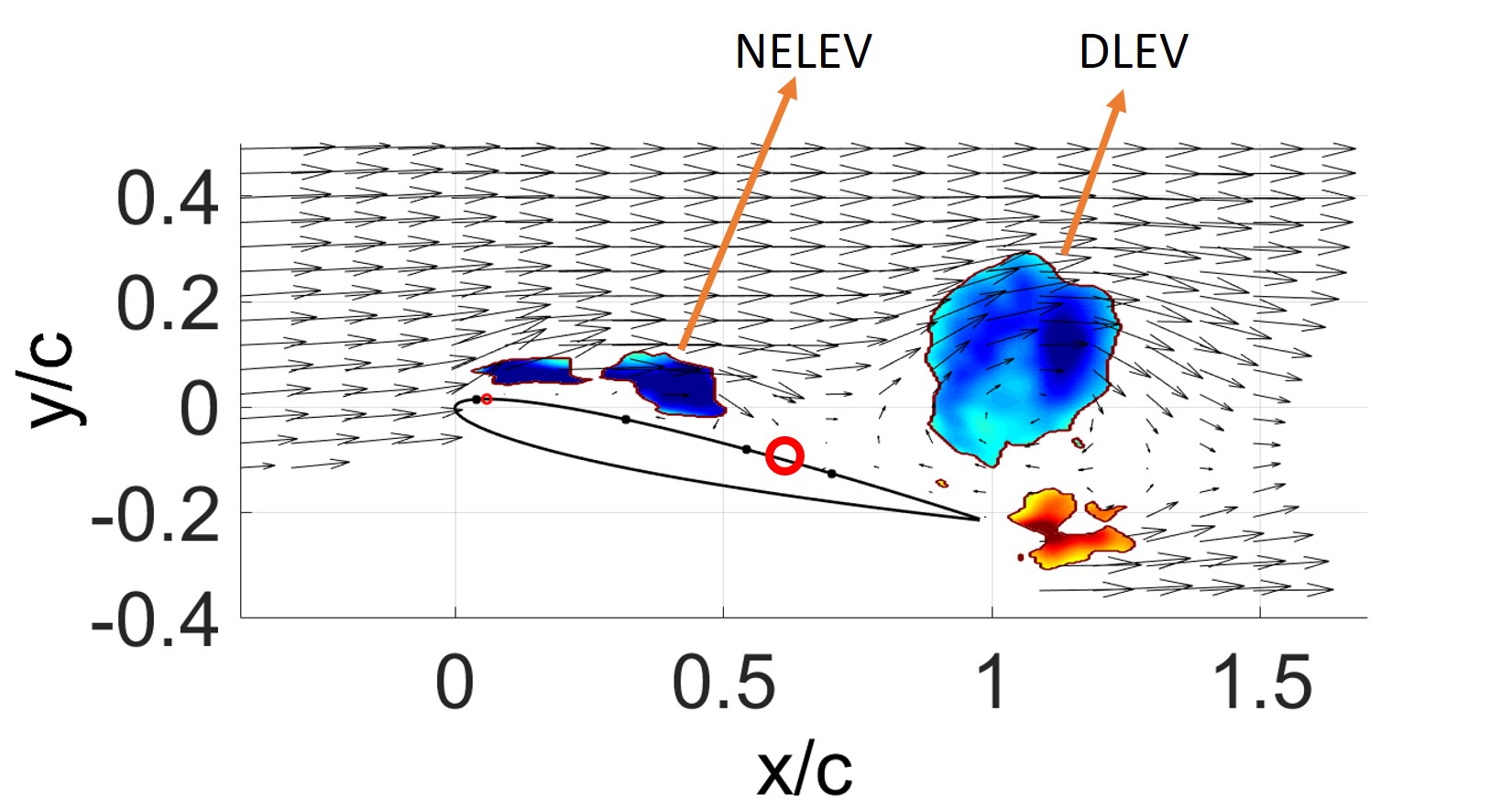}
	          \caption{$1.4t^+$}
	          \label{fig:Gamma2_1p4}
	\end{subfigure}	
	
	\begin{subfigure}{0.45\textwidth}
	        \includegraphics[width=2.8in]{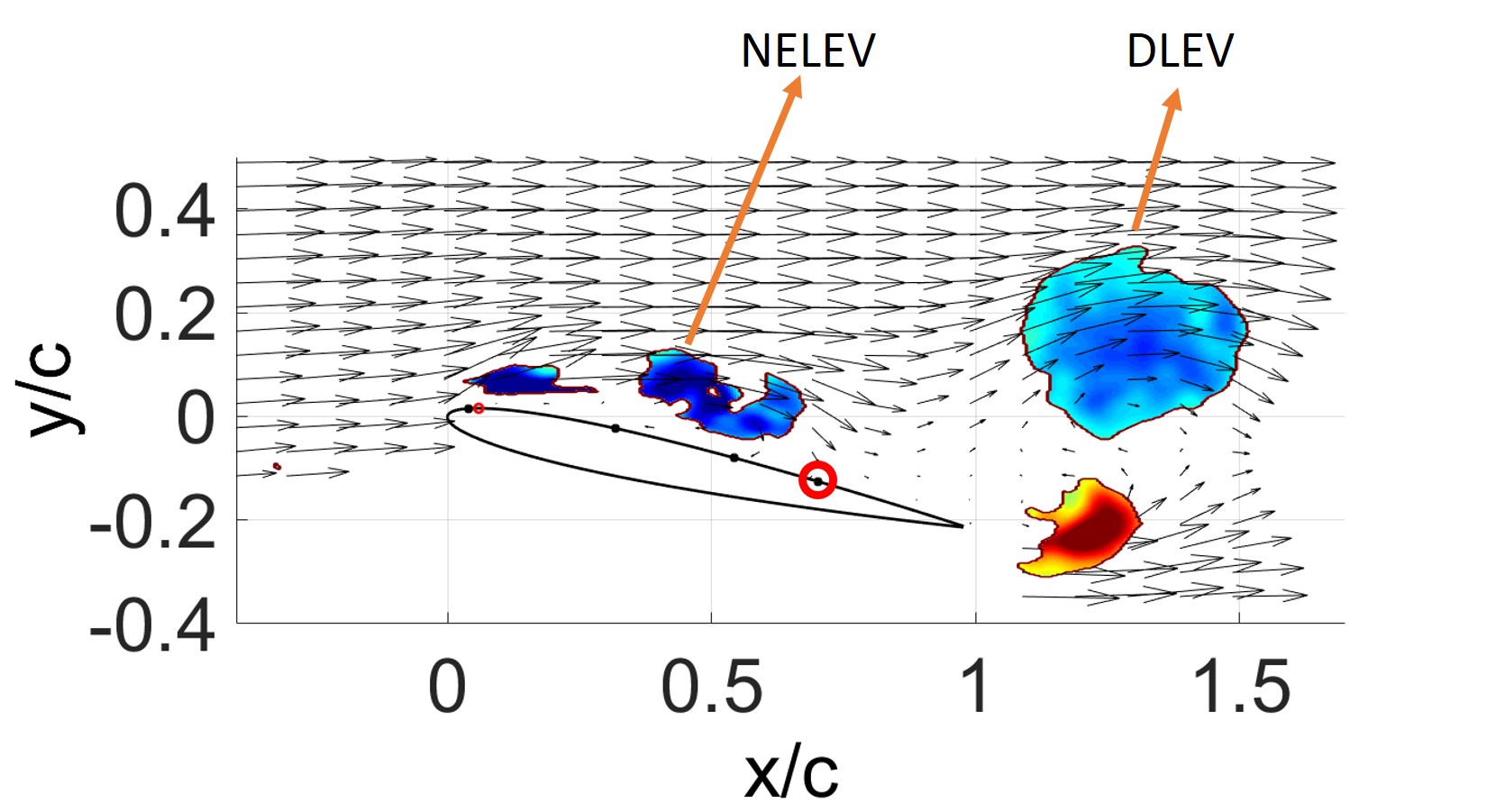}
	          \caption{$1.7t^+$}
	          \label{fig:Gamma2_1p7}
	\end{subfigure}
	
		\begin{subfigure}{0.45\textwidth}
		        \includegraphics[width=2.8in]{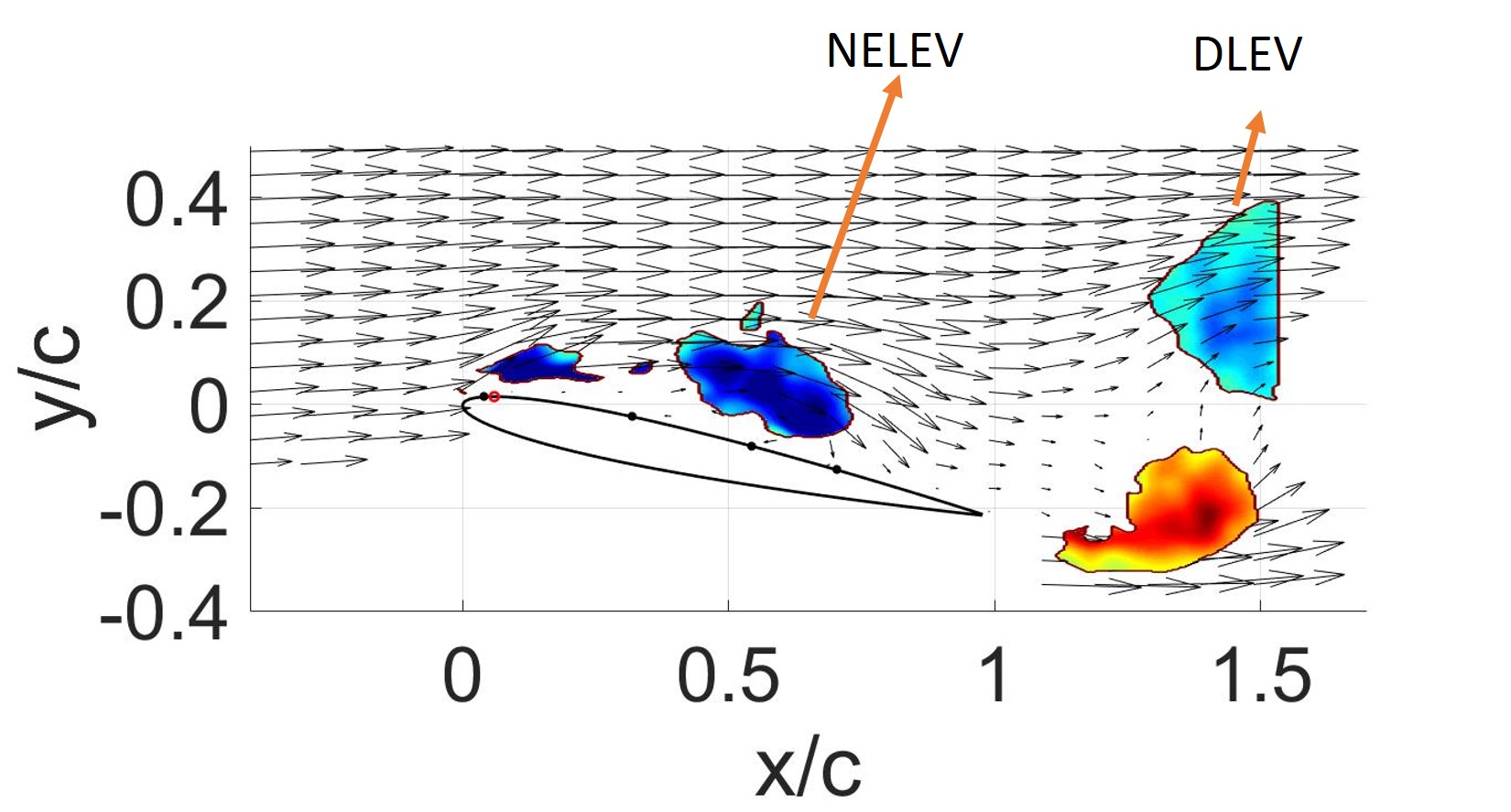}
		          \caption{$2t^+$}
		          \label{fig:Gamma2_2}
		\end{subfigure}
	
	\begin{subfigure}{0.45\textwidth}
	        \includegraphics[width=2.8in]{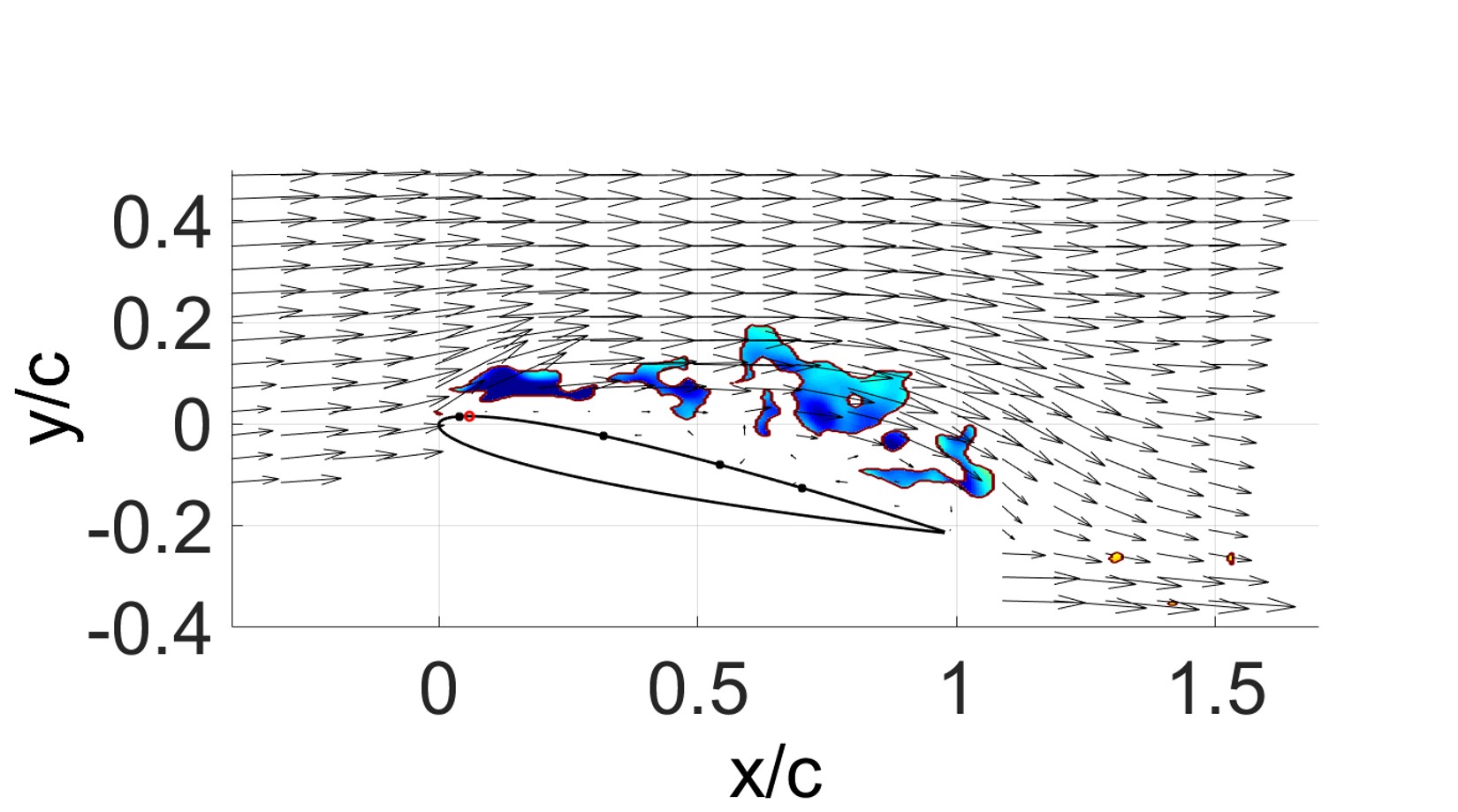}
	          \caption{$2.8t^+$}
	          \label{fig:Gamma2_2p8}
	\end{subfigure}		
	
	\begin{subfigure}{0.45\textwidth}
	        \includegraphics[width=2.8in]{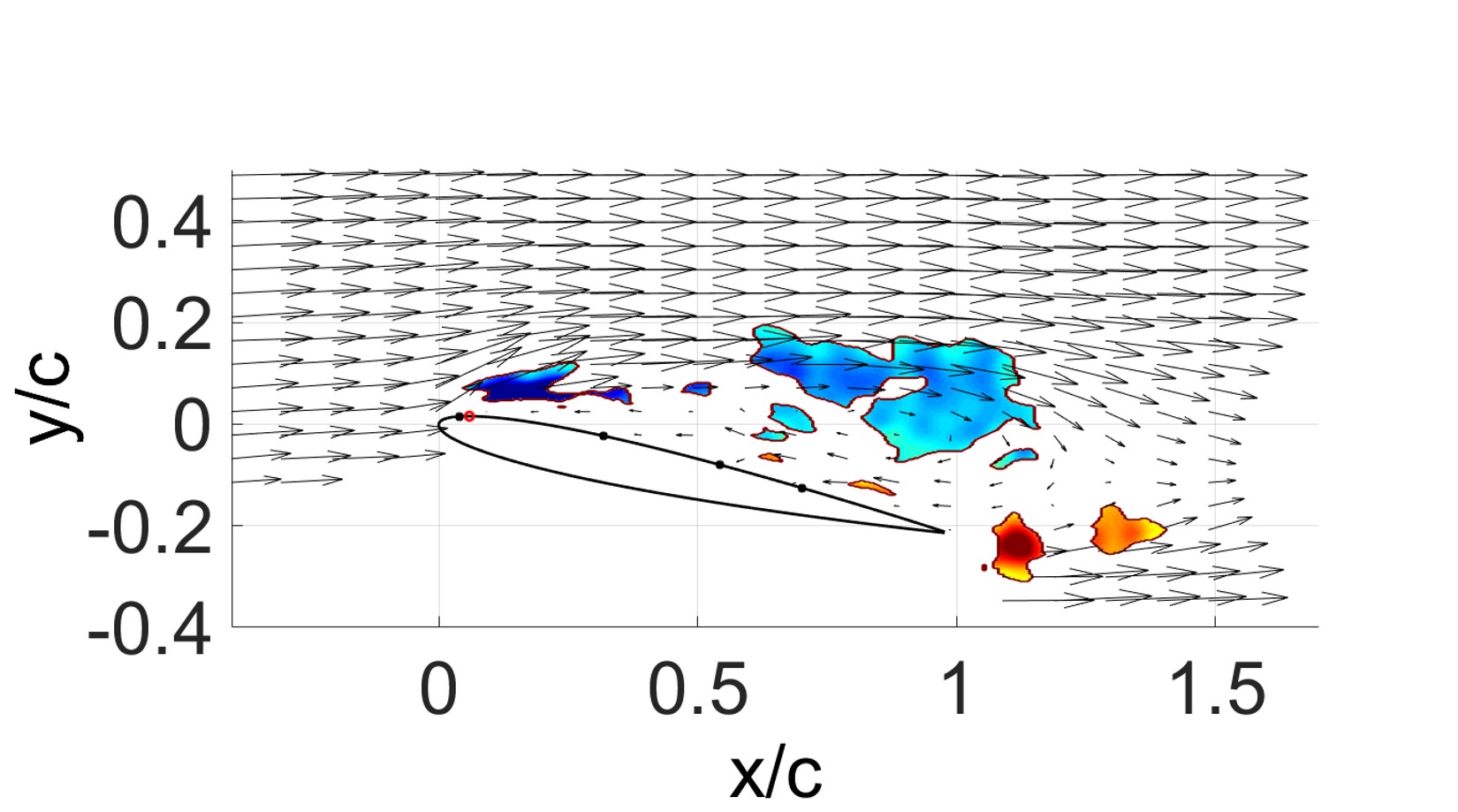}
		         \caption{$20t^+$}
              \label{fig:Gamma2_20}
	\end{subfigure}				


		

\end{multicols}

    \caption{Vorticity maps and velocity vectors following a single burst from the actuator. The red circles denote the location where the downwash impacts the wing. The red dot close to the leading-edge indicates the location of the actuators, and the black dots show the location of the pressure sensors. The plots from a to h represent baseline $(0t^+)$, maximum pressure on PS2 $(0.7t^+)$, maximum pressure on PS3 $(1.2t^+)$, minimum $C_L$ $(1.4t^+)$, maximum pressure on PS4 $(1.7t^+)$, vortices shed from the wing $(2t^+)$, maximum $C_L$ $(2.8t^+)$ and the baseline $(20t^+)$.}
    \label{fig:Gamma}		
\end{figure}

The vortex structures in figure \ref{fig:Gamma} are plotted at the 'critical' instants when $C_L$ reaches its maximum and minimum, the downwash (which will be discussed in more detail later) impinges on each pressure sensor as well as the instants right before the actuation is initiated and after the flow is fully relaxed. The vortex structure at $0t^+$ (figure \ref{fig:Gamma2_0}) shows a group of clockwise rotating vortices (blue) above the airfoil denoting the separated boundary layer. The counterclockwise rotating vortices (red) in the vicinity of the trailing edge indicate the trailing-edge vortex (TEV). As shown in figure \ref{fig:Gamma2_0p7}, the single-burst actuation excites the instability of the flowfield and the shear layer rolls up into two clockwise rotating vortices. The upstream part forms a new leading-edge vortex that bonds to the suction side of the airfoil. We will call this vortex the newly established leading-edge vortex or its abbreviation NELEV. The downstream portion of the shear layer rolls up into another vortex which eventually detaches from the airfoil. The previously discussed `kink' in the shear layer is at the interface between the two vortices.  The size of the downstream vortex grows continuously before it completely detaches from the airfoil (figure \ref{fig:Gamma2_1p2} to \ref{fig:Gamma2_1p7}). This vortex is referred to as the detached leading-edge vortex (DLEV). At $2.8t^+$ (figure \ref{fig:Gamma2_2p8}), both the DLEV and the counter-clockwise rotating TEV have detached from the airfoil. The flow becomes less separated than the baseline separated flow, and the resulting flow leaves the trailing edge smoothly. The reattachment point reaches the trailing edge at this time, and the lift increment reaches its maximum as previously shown in figure \ref{fig:flow_field}d and figure \ref{fig:single_CL_CM}d. Finally, as expected, at $20t^+$ after the actuation (figure \ref{fig:Gamma2_20}) the flow returns to the original baseline state.

The NELEV together with DLEV in figure \ref{fig:Gamma} produces a downward flow (downwash) that causes the pressure reversal at each pressure sensor following the burst actuation. The location, where the downwash impinges on the suction side surface of the wing is shown in figure \ref{fig:Gamma} b-e with red circles. At $0.7t^+$ after the firing of the burst (figure \ref{fig:Gamma2_0p7}), the downwash flow is impinging on PS2, when this pressure sensor is showing maximum pressure reading (figure \ref{fig:CL_CM_pressure}). Similarly, the peak in the pressure reversal occurs at PS3 and PS4 (figure \ref{fig:Gamma2_1p2} and figure \ref{fig:Gamma2_1p7}) when the downwash impinges on them.

\FloatBarrier

Therefore, by combining the results of the $\Delta C_L$ and $\Delta C_M$ measurements, the surface pressure measurements (figure \ref{fig:CL_CM_pressure} and figure \ref{fig:DCL_single_pressure}) and the vortex structure of the flowfield (figure \ref{fig:Gamma}), we conclude that the formation of the vortices, DLEV and NELEV produces a downwash that impinges on the suction side of the airfoil and convects downstream. This downwash contributes to a local pressure reversal and as a consequence, the sequence of local pressure reversal leads to the lift and pitching moment reversal following the single-burst actuation. 


\subsection{Proper Orthogonal Decomposition (POD) of the flow field}

Next, we apply the Proper Orthogonal Decomposition (POD) to the flowfield snapshots following the single-burst actuation to gain additional insight into the modes that are responsible for the lift and pitching moment reversal. 
This is inspired by the experiment conducted by \citet{monnier2016comparison}. They reported that the negative temporal coefficient of the second POD mode correlates with the lift coefficient variation following a single-burst actuation. 

The POD method can reduce a large number of interdependent variables to a much smaller number of independent modes while retaining as much as possible the variation in the original variables \citep{kerschen2005method}.

	\begin{equation} \label{eq:51}
	v(x,t)=\sum\limits_{i=1}^{\infty}a_i(t)\phi_i(x)
	\end{equation}	
in Eq. \ref{eq:51}, $a_i$ is the time-dependent temporal coefficient and $\phi_i(x)$ is the POD basis function.
Singular Value Decomposition (SVD) \citep{kerschen2005method} is performed on both horizontal and vertical velocity components obtained from the PIV measurements. 

For any given $(m\times n)$ matrix $X$
    \begin{equation} \label{eq:52}
    X=USV^T
    \end{equation}

where $U$ is an $(m\times m)$ orthonormal matrix containing the left singular vectors, $S$ is a $(m\times n)$ pseudo-diagonal and semi-positive definite matrix with diagonal elements $\delta_i$, and $V$ is an $(n\times n)$ orthonormal matrix containing the right singular vectors. There are physical meanings for each term from the SVD. The matrix $U$ represents the spatial distribution of velocity within each POD mode, $V$ contains the temporal coefficients for each mode, and the pseudo-diagonal elements of $S$ denote the energy level for the modes, in which the energy is descending with increasing of the mode number.        

\begin{figure}
	\centering
    \includegraphics[width=0.8\textwidth]{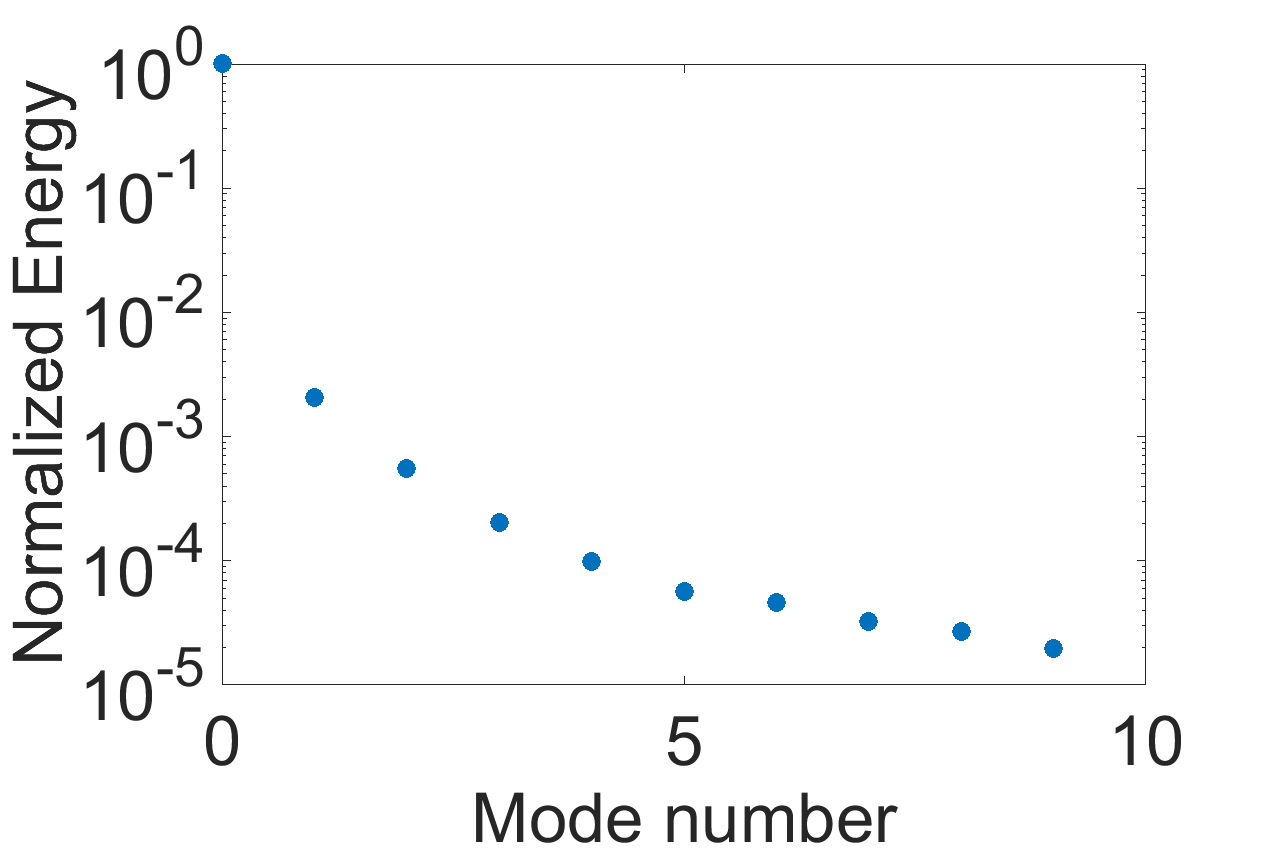}
    \caption{Energy distribution versus POD mode number.} 
    \label{fig:POD_energy} 
\end{figure}

The energy contained in each mode is normalized by mode 0 and plotted in figure \ref{fig:POD_energy}. It shows that about 84\% of the disturbed flow energy (with mode 0 subtracted from the flowfield) is contained in mode 1 and mode 2, which means that mode 1 and mode 2 can reconstruct a flowfield retaining most of the energetic structures in the disturbed flow.

\begin{figure}
	\centering

	\begin{subfigure}{0.45\textwidth}
	        \includegraphics[width=2.6in]{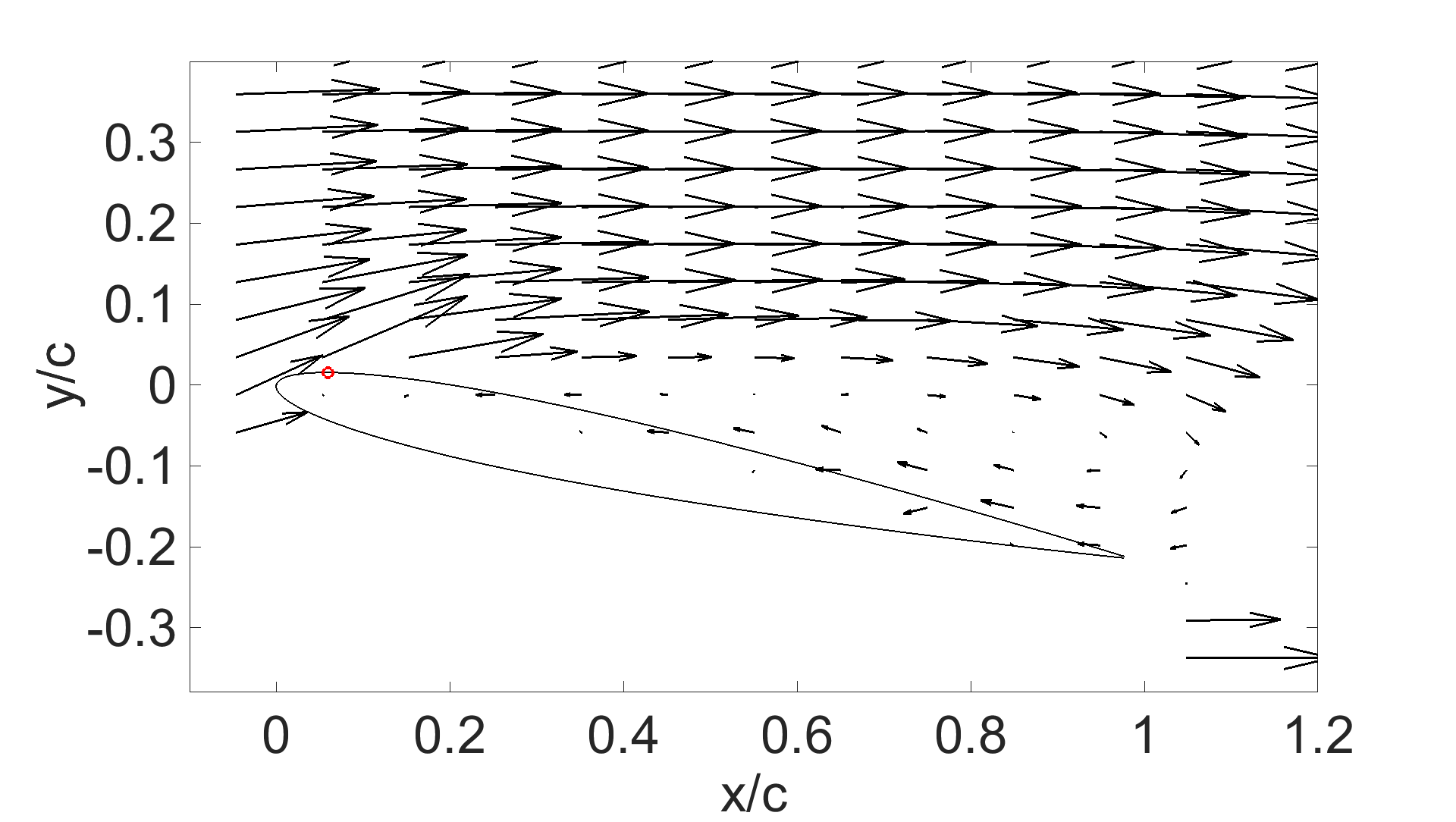}
			\caption{Mode 0}
	          \label{fig:POD_S_0}
	\end{subfigure}	
	~
	\begin{subfigure}{0.45\textwidth}
	        \includegraphics[width=2.6in]{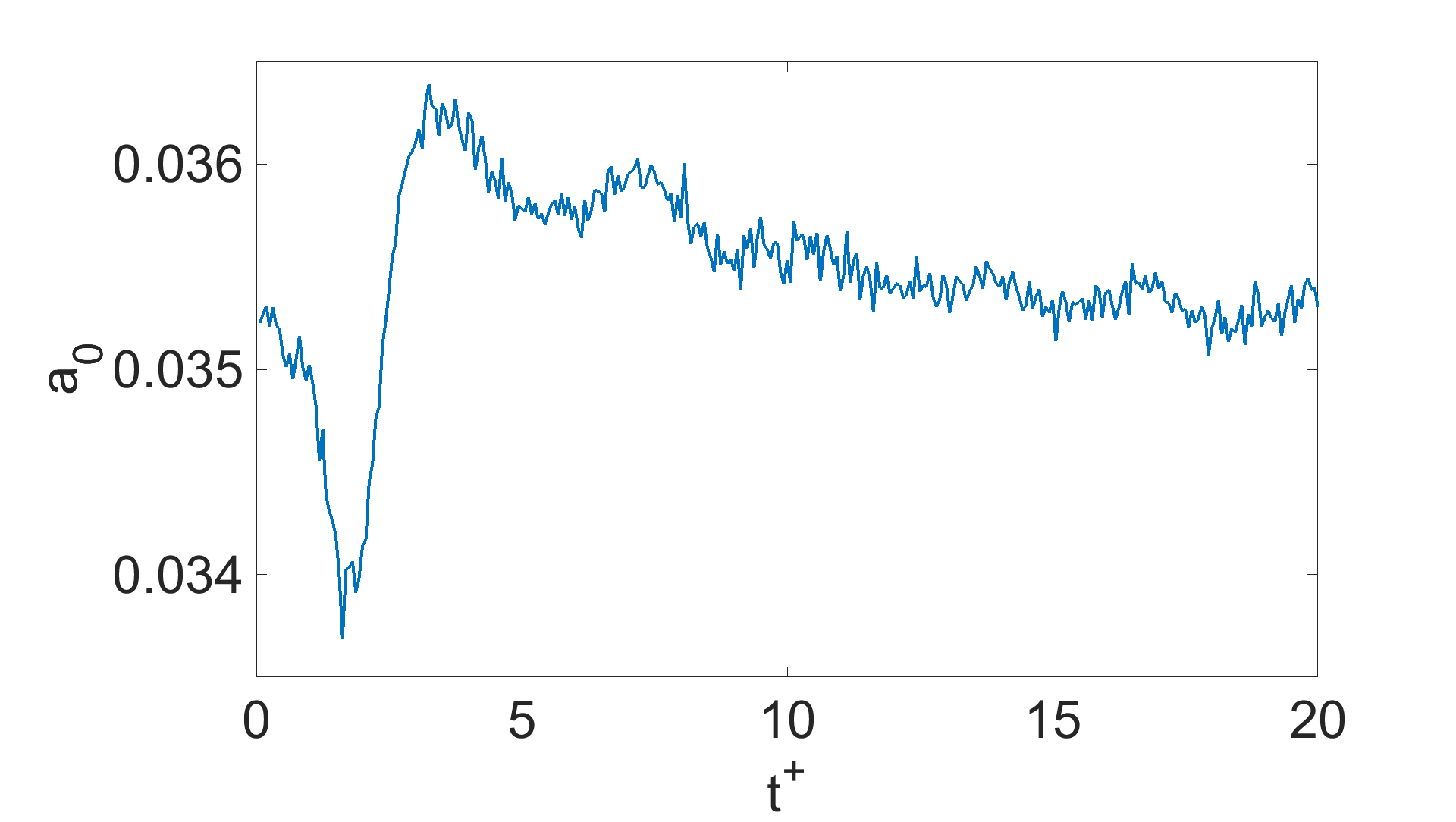}
	          \caption{Temporal coefficient of Mode 0}
	          \label{fig:POD_T_0}
	\end{subfigure}	
	
	\begin{subfigure}{0.45\textwidth}
	        \includegraphics[width=2.6in]{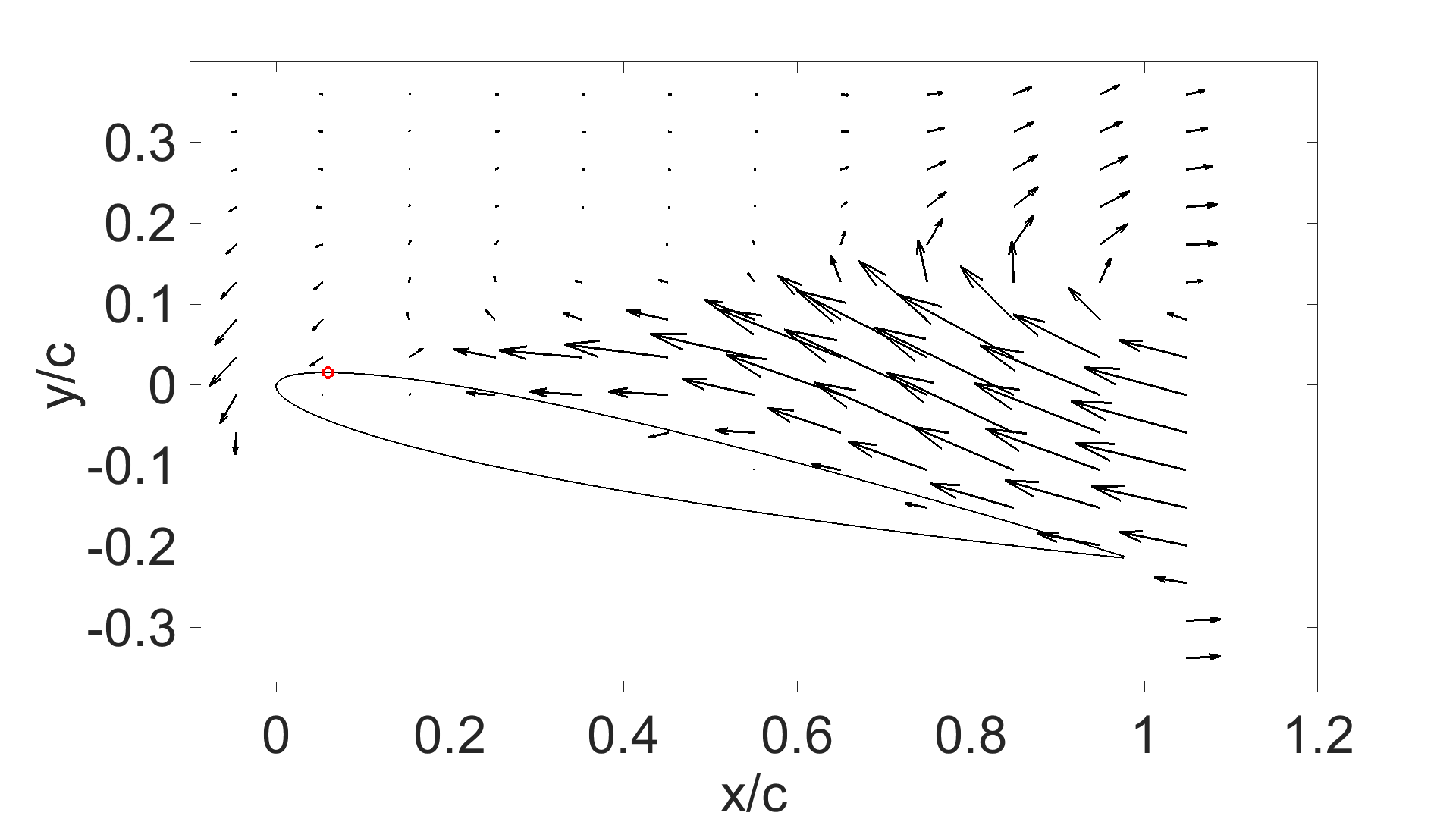}
			\caption{Mode 1}
	          \label{fig:POD_S_1}
	\end{subfigure}	
	~
	\begin{subfigure}{0.45\textwidth}
	        \includegraphics[width=2.6in]{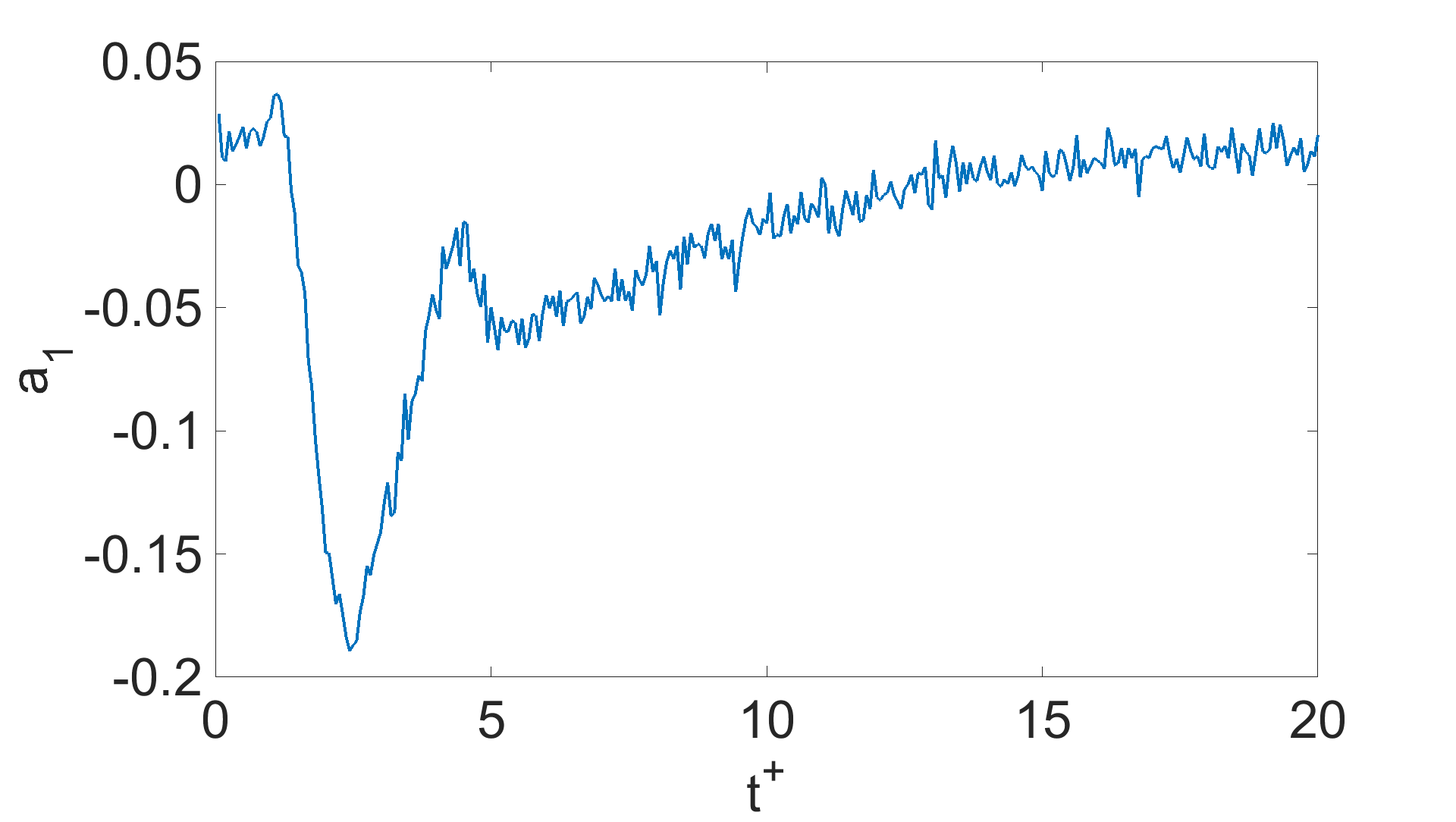}
	          \caption{Temporal coefficient of Mode 1}
	          \label{fig:POD_T_1}
	\end{subfigure}	

	\begin{subfigure}{0.45\textwidth}
	        \includegraphics[width=2.6in]{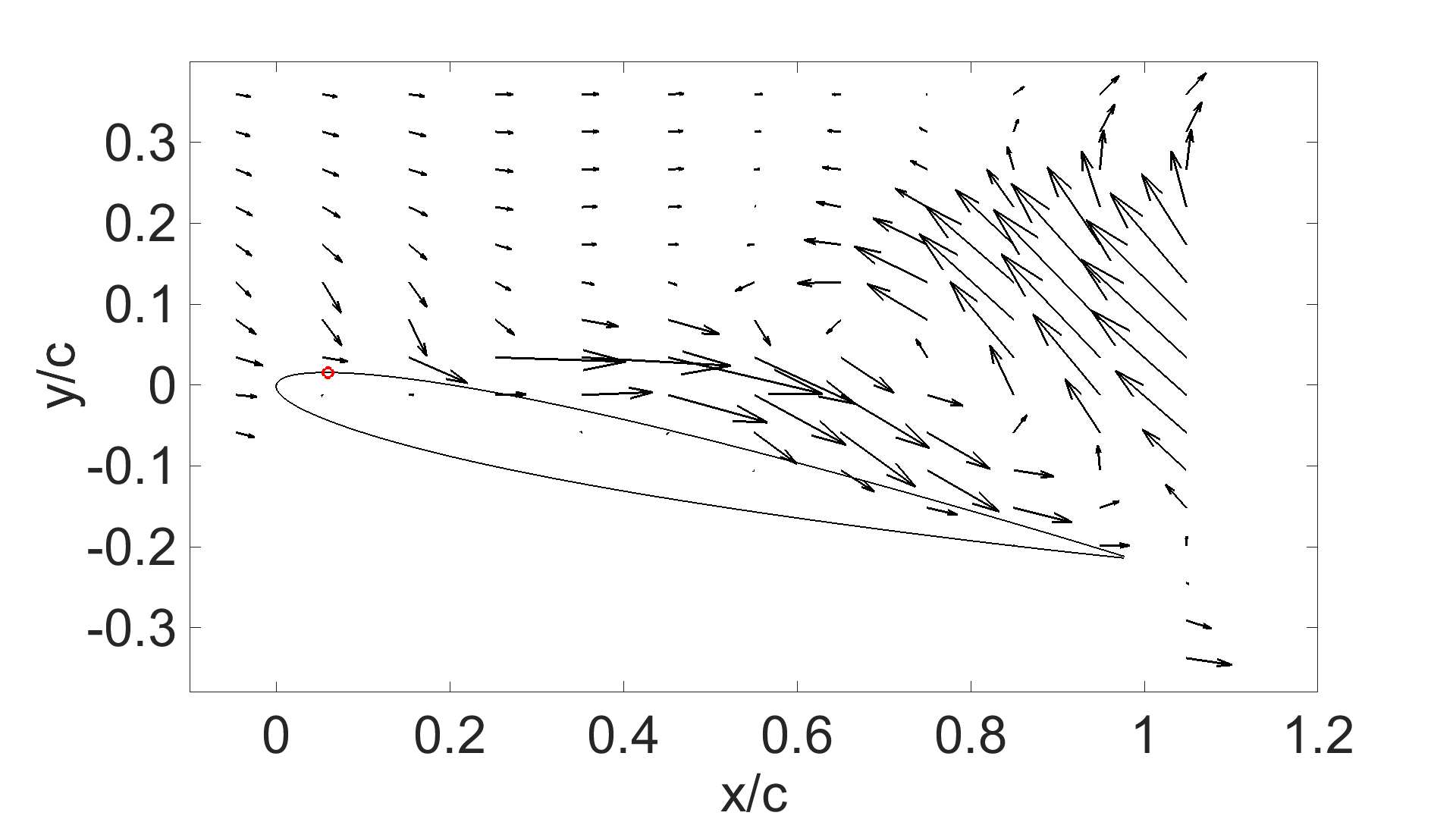}
			\caption{Mode 2}
	          \label{fig:POD_S_2}
	\end{subfigure}	
	~
	\begin{subfigure}{0.45\textwidth}
	        \includegraphics[width=2.6in]{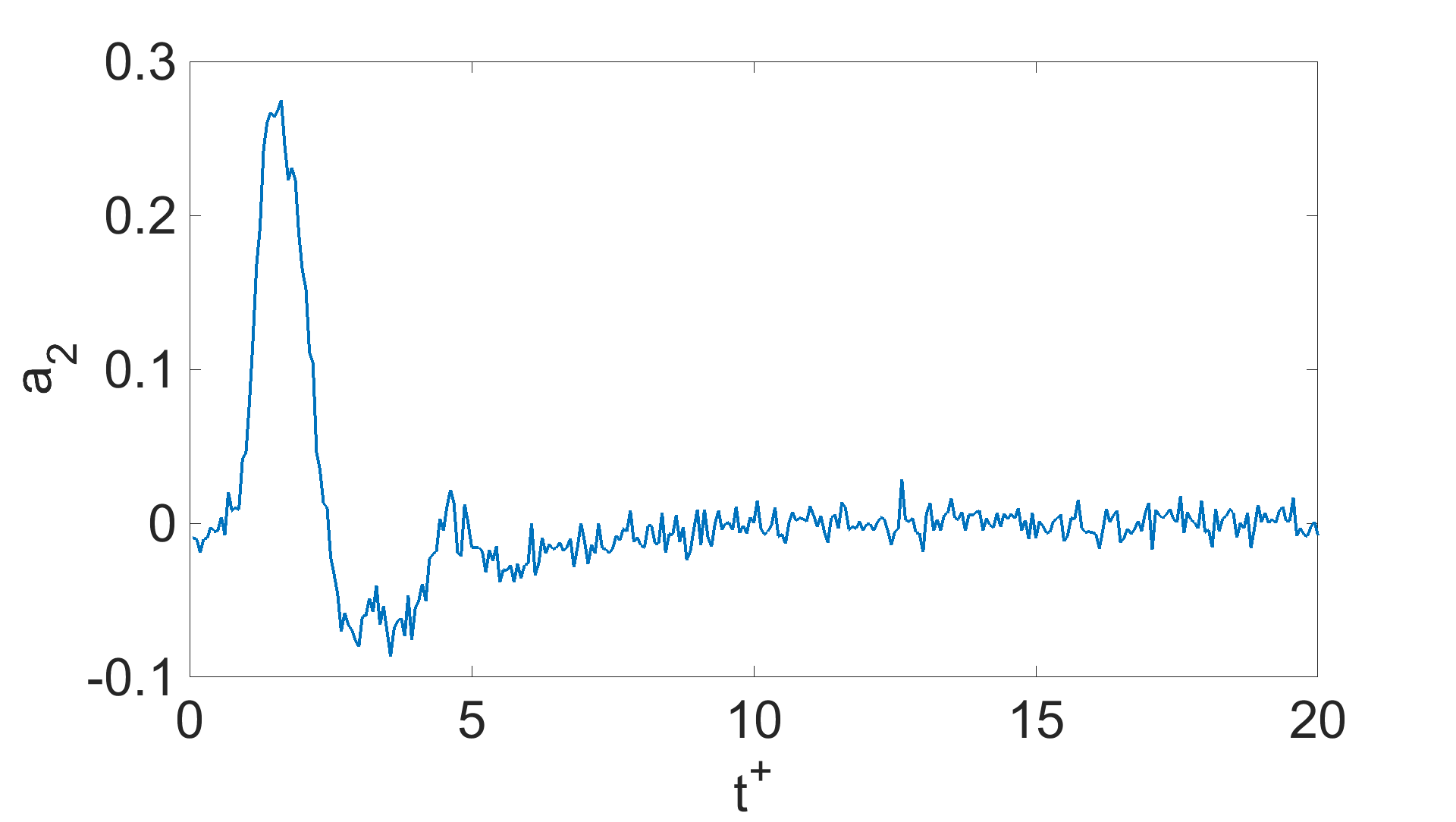}
	          \caption{Temporal coefficient of Mode 2}
	          \label{fig:POD_T_2}
	\end{subfigure}

				
   	   	
    \caption{Comparison of the first four spatial POD modes and their corresponding temporal coefficients.}
    \label{fig:POD}		
\end{figure}

The corresponding basis functions and the temporal coefficients for the first three modes are plotted in figure \ref{fig:POD}. Mode 0 (figure \ref{fig:POD_S_0}) shows the flow structure that is very similar to the baseline separated flow. Figure \ref{fig:POD_T_0} shows that mode 0 almost remains unchanged with time. The temporal coefficient of mode 0 is always above 0 which means the directions of the velocity vectors in figure \ref{fig:POD_S_0} are preserved. Mode 1 shown in figure \ref{fig:POD_S_1} is related to the reverse flow along the airfoil that causes flow separation. The temporal coefficient (figure \ref{fig:POD_T_1}) of this mode goes negative $1.2 t^+$ after the burst is initiated, which implies that the velocity vectors in this spatial mode flip their directions and thus, contribute to flow reattachment thereafter until its fully relaxed. Mode 2 shown in figure \ref{fig:POD_S_2} represents the DLEV downstream of the trailing edge, NELEV on the surface of the airfoil and the associated downwash impinging on the upper surface of the airfoil at the vicinity of PS4. The temporal coefficient of this mode (figure \ref{fig:POD_T_2}) goes above 0 following the burst until $2.5t^+$, which preserves the directions of the velocity vectors in figure \ref{fig:POD_S_2}. This implies that mode 2 is related to the $C_L$ and $C_M$ reversal. Next, we will make a direct comparison between the $C_L$, $C_M$ and the POD modes.      


 \FloatBarrier

\FloatBarrier

\begin{figure}
	\centering
	
		\begin{subfigure}{0.45\textwidth}
		        \includegraphics[width=2.6in]{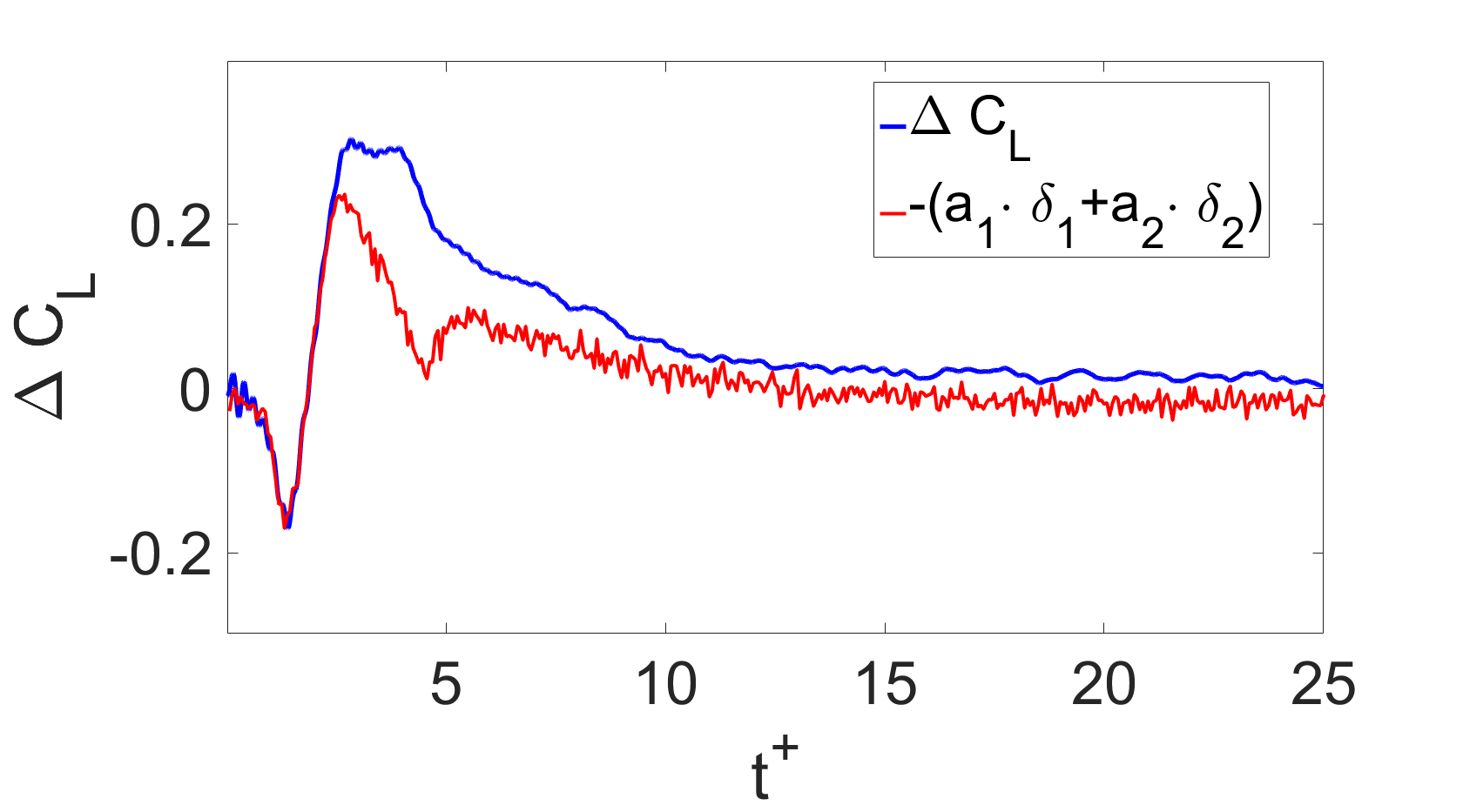}
		          \caption{Sing-burst actuation $\Delta C_L$}
		          \label{fig:DCL_POD}
		\end{subfigure}
~		
		\begin{subfigure}{0.45\textwidth}
		        \includegraphics[width=2.6in]{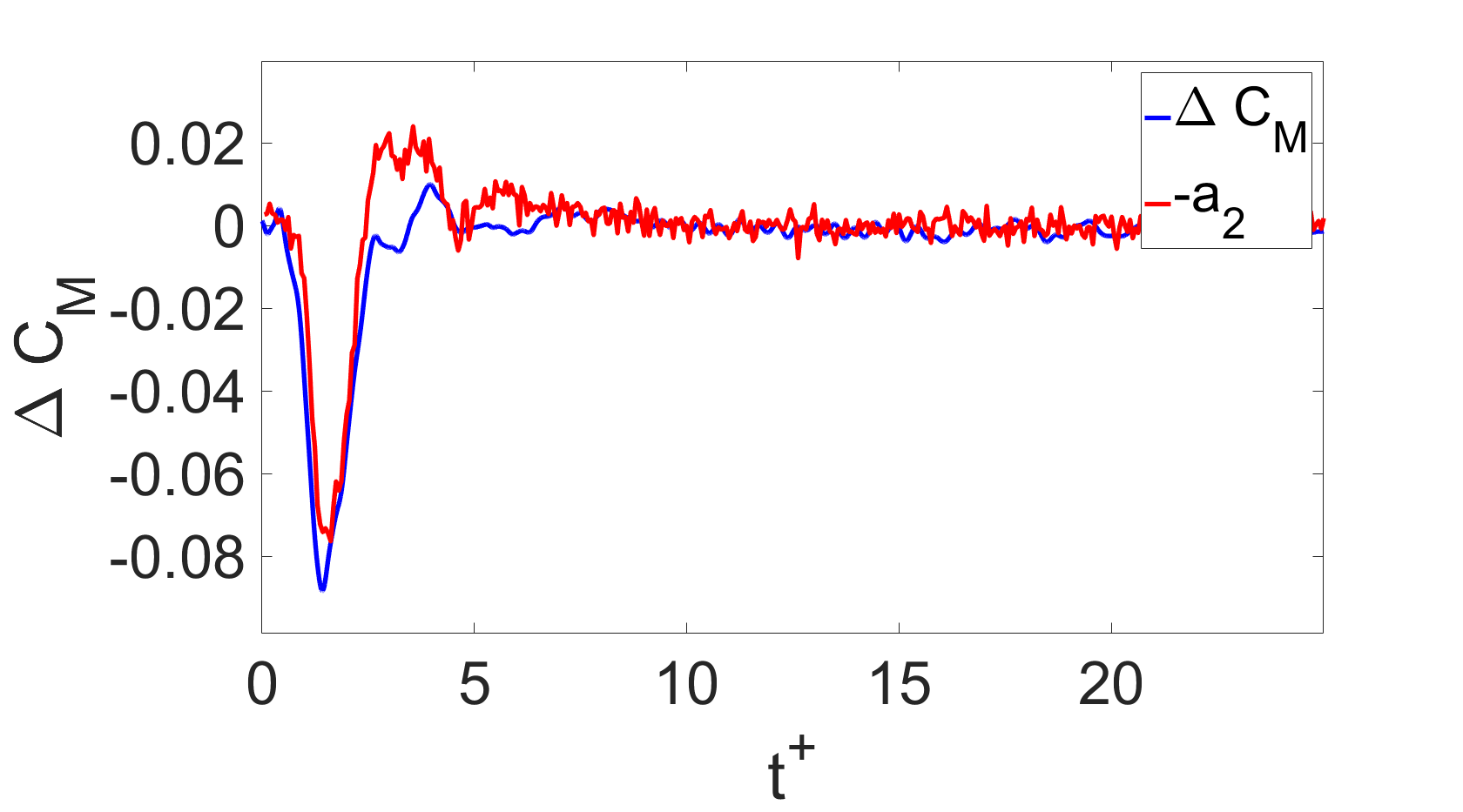}
		          \caption{Single-burst actuation $\Delta C_M$}
		          \label{fig:DCM_POD}
		\end{subfigure}


    \caption{Comparison of aerodynamic loads and time varying coefficient of POD modes, (a) lift coefficient, (b) pitching moment coefficient.}
    \label{fig:CL_CM_POD}		
\end{figure}

The sum of the negative temporal coefficients of mode 1 and mode 2 is plotted in figure \ref{fig:CL_CM_POD}a. Note that all the time coefficients of the POD modes are normalized by the minimum of the $\Delta C_L$ or $\Delta C_M$ respectively. Comparing $\Delta C_L$ to the combination of mode 1 and mode 2, $a_1\cdot \delta_1+a_2\cdot \delta_2$, the negative temporal coefficient of mode 1 and mode 2 combination tracks the $\Delta C_L$ very well (especially for the lift reversal), other than mode 2 alone, reported by \citet{monnier2016comparison}. 

Meanwhile, the negative temporal coefficient of mode 2 alone is closely tracking the measured $\Delta C_M$, since this mode represents the downwash impinging on PS4 (\ref{fig:CL_CM_POD}b), which has the longest momentum arm from the reference point. This observation again, suggests that the pitching moment reversal is caused by the downwash impinging on PS4 that is represented by POD mode 2. 

\FloatBarrier

\begin{figure}
	\centering
    \includegraphics[width=0.8\textwidth]{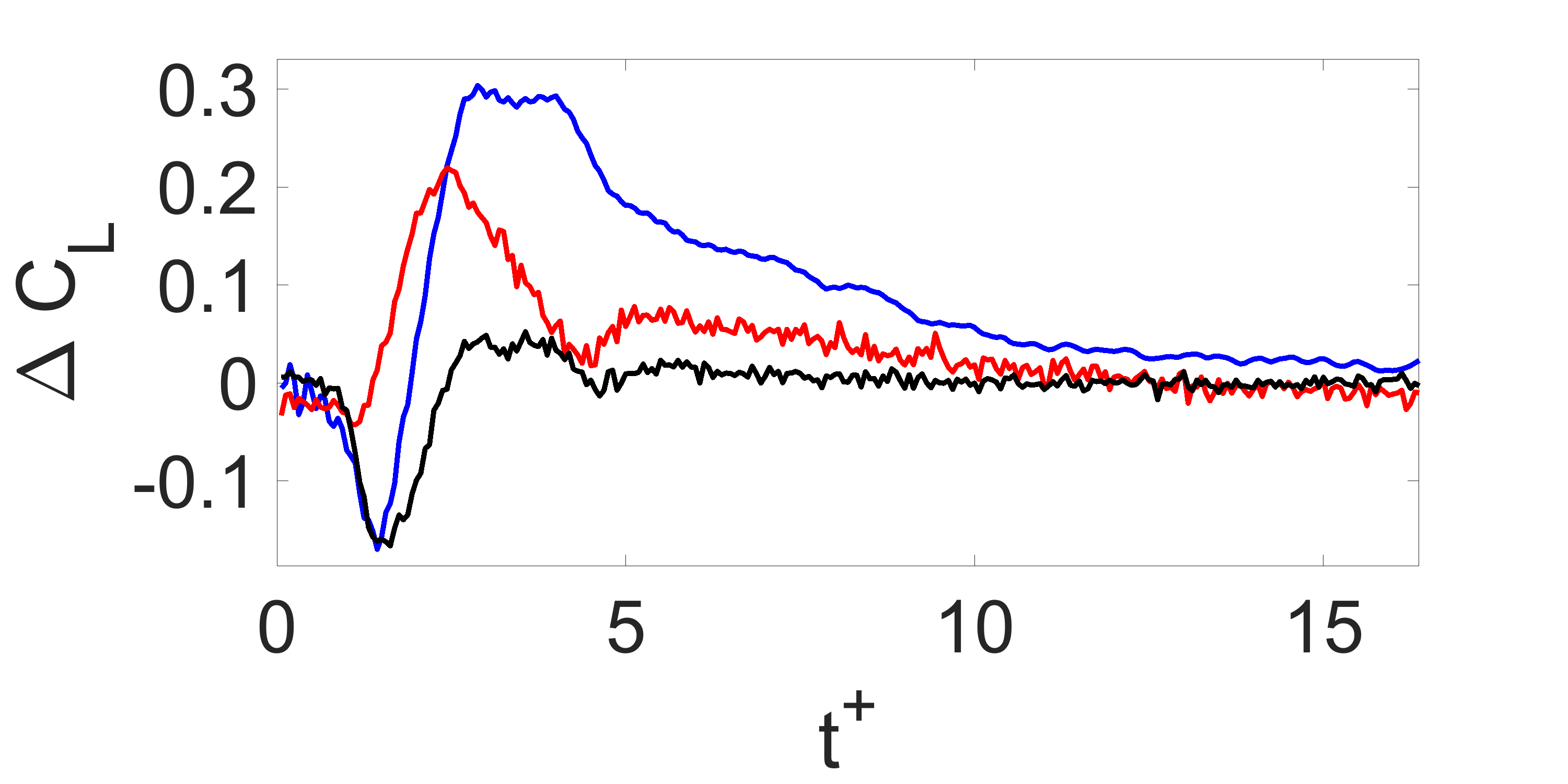}
    \caption{POD mode 1, POD mode 2 and $\Delta C_L$.} 
    \label{fig:POD_1_2} 
\end{figure}

On the other hand, since the lift reversal is tracked by the combination of mode 1 and mode 2, another direct comparison between each individual POD mode and $\Delta C_L$ is carried out in figure \ref{fig:POD_1_2}. It shows that the negative time coefficient of mode 2 follows $\Delta C_L$ from the initialization of the burst to $1.2t^+$ while the time coefficient of mode 1 almost remains constant with a positive value (which indicates a reverse flow on the airfoil or in other words, flow is separated). During this time period ($0t^+$ to $1.2t^+$), the lift reversal is perfectly captured by mode 2. After $1.2t^+$, mode 2 starts to deviate from $\Delta C_L$ only because the time coefficient of mode 1 starts to vary and goes up. Therefore, the good correlation between the negative coefficient of POD mode 2 and the lift reversal curve suggests that this mode is responsible for the lift reversal. In other words POD mode 2 which represents the formation of NELEV and DLEV associated with their induced downwash cause the lift reversal. In addition, the time coefficient of mode 2 is almost always equal to zero except when the lift/pitching moment reversal occurs, which suggests that this flow pattern contributes to the lift/pitching moment reversal only. On the other hand, the comparison between POD mode 1 and $\Delta C_L$ in figure \ref{fig:POD_1_2} suggests POD mode 1, which represents the reverse flow on the surface of the airfoil contributes to the lift enhancement.

 \FloatBarrier


\section{Conclusion}\label{Sec:conc} In this work, the mechanism of the lift and pitching moment reversal following a single-burst actuation on a stalled NACA-0009 airfoil is studied experimentally. The time-evolving flow structure, lift, pitching moment, pressure and their dynamic characteristics are analyzed. The maximum lift and pitching moment reversal is observed at the vicinity of $1.4t^+$ after the initiation of the single-burst, and the maximum lift increment occurs at $2.8t^+$, when the flow is reattached. We observed that
following the initiation of the actuation, there is a downwash impinging on the suction side of the wing that moves with the clockwise rotating large-scale leading-edge vortices. This results in a pressure reversal and thus, leads to the lift and pitching moment reversal before the DLEV convects into the wake. A further POD analysis shows that 84\% of the disturbed kinetic energy is retained in mode 1 and mode 2. The combined negative temporal coefficients of POD mode 1 and mode 2 track the lift coefficient curve well, especially during the lift reversal. Comparing the time coefficient of POD mode 2 with the lift and pitching moment variation, it confirms that the flow structure representing the downwash induced by DLEV and NELEV is responsible for lift and pitching moment reversal. We also found POD mode 1 which represents the strength and direction of the reverse flow on the suction side of the airfoil is responsible for the lift enhancement. The outcome of this paper could potentially benefit the fluidic actuator design to reduce the lift and pitching moment reversal and thus, leads to faster actuation response for unsteady flow separation control.

\bibliographystyle{jfm}
\bibliography{jfm}

\end{document}